\documentclass[A4,10pt]{iopart}
\usepackage{amsfonts}
\usepackage{amssymb,amsbsy}
\usepackage{geometry}
\usepackage{graphicx}
\usepackage{dsfont}

\begin{document}

\title{Remarks about the thermodynamics of astrophysical systems in mutual interaction and related notions}
\author{L Velazquez}
\address{Departamento de F\'\i sica, Universidad Cat\'olica del Norte, Av. Angamos 0610, Antofagasta,
Chile.}
\ead{lvelazquez@ucn.cl}

\begin{abstract}
General aspects about the thermodynamics of astrophysical systems are discussed, overall, those concerning to astrophysical systems in mutual interaction (or the called \emph{open astrophysical systems}).  A special interest is devoted along the paper to clarify several misconceptions that are still common in the recent literature, such as the direct application to the astrophysical scenario of notions and theoretical frameworks that were originally conceived to deal with extensive systems of the everyday practice (large systems with short-range interactions). This discussion starts reviewing the current understanding about the notion of \emph{negative heat capacity}. Beyond to clarify its physical relevance, it is discussed the conciliation of this notion within classical fluctuation theory, as well as equilibrium conditions concerning to systems with negative heat capacities. These results motivate a revision of our understanding about critical phenomena, phase transitions and the called zeroth law of thermodynamics. Afterwards, general features about the thermodynamics of astrophysical systems are presented throughout the consideration of simple models available in the literature. A particular attention is devoted to the influence of evaporation on the macroscopic behavior of these systems. These antecedents are then applied to a critical approach towards the thermodynamics of astrophysical systems in mutual interaction. It is discussed that the long-range character of gravitation leads to the incidence of long-range correlations. This peculiarity imposes a series of important consequences, such as the non-separability of a single astrophysical structure into independent subsystems, the breakdown of additivity and conventional thermodynamic limit, a great sensibility of the macroscopic behavior to the external conditions, the restricted applicability of the called thermal contact in astrophysics, and hence, the non-relevance of conventional statistical ensembles in this scenario. To clarify how some of conventional notions and theoretical frameworks could be extended to open astrophysical systems, an exploratory study of a paradigmatic situation is presented: \emph{a binary astrophysical system}. This analysis is carried out in the framework of the \emph{quadrupole approximation}, which represents the lowest coupling among internal and collective degrees of freedom. Apparently, collective motions are responsible of a non-linear energy interchange among the astrophysical systems. This mechanism introduces some modifications in equilibrium conditions for stationary and stability, such as a generalization of Thirring's stability condition for systems with negative heat capacities [Z. Phys. \textbf{235}, 339 (1970)]. Additionally, the stability of collective motions of this binary astrophysical system is also discussed, which is related to the low energy thermodynamic behavior of the model discussed by Votyakov and co-workers [Phys. Rev. Lett. \textbf{89}, 031101 (2002)]. The thermodynamic limit for self-gravitating gas of identical non-relativistic point particles is then derived and compared with other different proposals. The astrophysical counterpart of Gibbs-Duhem relation is obtained and compared with recent proposal of Latella and co-workers [Phys. Rev. Lett. \textbf{114}, 230601 (2015)]. Finally, it is analyzed the incidence of nonextensivity during the merge of two identical astrophysical systems. On the contrary to the situation considered in \emph{Gibbs paradox}, the merge is an irreversible process that crucially depends on the existence (or nonexistence) of the external gravitational influence of other systems.
\end{abstract}

\tableofcontents
\newpage

\section{Introduction}

Astrophysical structures observed in the Universe, including the Universe itself, pose challenging questions for the conventional thermodynamics and statistical mechanics. Gravitation is a long-range interaction that does not satisfy the \emph{stability and regularity properties} \cite{Reichl,Gallavotti,Binney,Bertin}, which are basic requirements for a self-consistent statistical mechanical description and the validity of theorems assuring the equivalence of statistical ensembles in the thermodynamic limit. Considerable efforts have been devoted in the last decades for understanding thermo-statistical properties of long-range interacting systems, in particular, the astrophysical systems \cite{Dauxois1,Dauxois2,Gross,Campa.1,Campa.2}.

The motivation for this paper born from the reading of some recent contributions on the thermo-statistics of gravitational clustering phenomenon in Cosmology  \cite{SaslawHamilton,SaslawB,Saslaw,Itoh}. As expected, thermodynamic description provides a different view of this phenomenon beyond the standard picture obtained from observations and $N$-body cosmological simulations. Many works available in the literature start from the consideration of conventional ensembles of statistical mechanics such as canonical or gran-canonical ensembles \cite{SaslawHamilton,Saslaw}. These studies typically involve the calculation of partition function:
\begin{equation}\label{ZZ}
Z_{N}(T,V)=\frac{1}{\Lambda^{N}N!}\int\exp\left[-\beta H_{N}(\mathbf{r},\mathbf{p})\right]d^{N}\mathbf{r}d^{N}\mathbf{p}
\end{equation}
associated with a self-gravitating system of $N$ non-relativistic particles with Hamiltonian $H_{N}(\mathbf{r},\mathbf{p})$. Partition function depends on the system volume $V$ and its average temperature $T$, which enters into expression (\ref{ZZ}) via the control parameter $\beta=1/kT$. The constant $\Lambda$ normalizes the phase space volume cell. This type of statistical description, however, is not fully justified according to the present understanding about thermodynamics of astrophysical systems.

This work is devoted to discuss limitations of conventional thermo-statistical treatments in their application to astrophysical situations, as well as the revision of some thermodynamic notions concerning to systems in mutual interaction. The paper is organized into sections as follows. The second section is devoted to review the current understanding of the negative heat capacity phenomenon, which is a distinguishing macroscopic behavior in the thermodynamics of astrophysical systems. A special attention is paid to the stability conditions and the fluctuation theorems compatible with systems with negative heat capacities. These thermo-statistical results enable us to perform a re-examination of several important notions in statistical mechanics. The third section is devoted to review some general thermodynamic features of astrophysical systems throughout the consideration of simple models proposed in the literature. The main contributions of this work are presented in fourth section. Special interest will be devoted to the revision of the notions of separability and additivity, as well as their relevance in the thermodynamic stability conditions that are applicable to astrophysical systems in mutual interaction. This analysis will be focussed on the consideration of a binary astrophysical system, a simple but paradigmatic situation to clarify the role of the called \emph{quadrupole-orbit coupling} and the angular momentum conservation on the stability of astrophysical systems in mutual interaction. The question about the non-extensivity of astrophysical systems will be also addressed, as well as the possible relevance of a thermodynamic limit for the case of self-gravitating gas of identical non-relativistic point particles. Finally, some conclusions and open problems are summarized in the fifth section.

\section{Negative heat capacities: relevance, misconceptions and implications}\label{NHC}

\subsection{Negative heat capacity: A conflicting thermodynamic notion}

Gravitational clustering is a sudden process of re-organization that leads to development of compact astrophysical structures with negative heat capacities. Although this thermodynamic behavior is regarded as an anomaly from the conventional perspective of statistical mechanics and thermodynamics, its presence is more a rule rather than an exception in the framework of astrophysical systems. In fact, this behavior plays a fundamental role during their evolution. It explains, as example, how interior of stars become hotter while shrink under its own gravitational field, a process that enables the occurrence of nuclear fusion reactions among more heavy atomic nuclei along the stellar evolution. The ubiquitous character of negative heat capacities in astrophysics can be justified through simple arguments as the virial theorem \cite{Eddington}, or even more elaborated astrophysical models such as Antonov isothermal model \cite{Antonov} and King models of star clusters \cite{King}.
Nowadays, the existence of negative heat capacities has been reported in many other physical scenarios \cite{Padmanabhan,Moretto,Lynden-Bell3,DAgostino,Ison,Einarsson} and its physical relevance is now widely accepted. Nevertheless, there still survive several misconceptions and open questions concerning to this notion. For example, there is a poorly knowledge in physical community about thermodynamical equilibrium conditions associated with negative heat capacities and their unexpected implications on conventional ideas of statistical mechanics and thermodynamics.

\subsection{Explanation of negative heat capacities paradox}

Eddington was ones of the first investigators to advertise that a star or a stars cluster expands and cool down by absorbing energy \cite{Eddington}. His arguments were based on virial theorem of classical mechanics \cite{Goldstein}. Decades later \cite{Lynden-Bell1}, Lynden-Bell and Wood considered the concept of \emph{negative specific heat} to explain gravothermal catastrophe of isothermal model of Antonov \cite{Antonov}. While astronomers enjoyed Eddington incongruous conclusions, the notion of negative specific heat (or negative heat capacity)  was received that time with skepticism by the statistical mechanics community \cite{Lynden-Bell3,Lynden-Bell2}. As already emphasized two years later by Thirring \cite{Thirring}, the known theorem of classical fluctuation theory \cite{Landau}:
\begin{equation}\label{can.fr}
C=k\beta^{2}\left\langle\delta U^{2}\right\rangle
\end{equation}
seems to suggest that heat capacity should be \emph{nonnegative}. Thirring himself explained this paradox through the \emph{inequivalence} of canonical and microcanonical statistical ensembles. Fluctuation relation (\ref{can.fr}) can be straightforwardly obtained by combining the first derivatives of canonical partition function (\ref{ZZ}):
\begin{equation}\label{partition}
\left\langle U\right\rangle=-\frac{\partial}{\partial\beta}k\log Z(\beta)\mbox{ and }\left\langle \delta U^{2}\right\rangle=-\frac{\partial^{2}}{\partial\beta^{2}}k\log Z(\beta)
\end{equation}
where $\beta=1/kT$, with $T$ being environmental temperature (the system is found in thermal equilibrium with a heat bath or a thermostat), and $k$ is Boltzmann constant. However, a different perspective is obtained when one starts from microcanonical definition of temperature:
\begin{equation}\label{mic.def}
\frac{1}{T}=\frac{\partial S}{\partial U}\rightarrow-\frac{1}{T^{2}}\frac{\partial T}{\partial U}=\frac{\partial^{2} S}{\partial U^{2}}\Rightarrow C=-\frac{\left(\partial S/\partial U\right)^{2}}{\left(\partial^{2} S/\partial U^{2}\right)}.
\end{equation}
Accordingly, a closed system exhibits a negative heat capacity when the entropy $S$ is a \emph{convex function} on the energy. As extensively discussed by Gross \cite{Gross}, this mathematical behavior arises in nonhomogeneous systems, such as the long-range interacting systems like the astrophysical ones, or finite systems with short-range interactions that undergo a phase coexistence phenomenon, etc. Thermodynamically speaking, the heat capacity is a \emph{response function}. If the system macroscopic state is determined by several control variables $X$ besides energy, the second order derivatives of entropy can be related to other response functions such as isothermal magnetic and electric susceptibilities, the isothermal compressibility, etc. Consequently, the existence of negative values in the heat capacity is just a particular case of anomalous values in response functions, whose mathematical origin is also related to the existence of states where the entropy is a non-concave function.

\subsection{Thirring stability conditions for systems with negative heat capacities}

A general misunderstanding concerning to states with negative heat capacities is that the presence of this anomaly, by itself, implies thermodynamical instability of these states. Although it could sound strange, \emph{thermodynamical stability of a given system is not an intrinsic property of the same one, but it crucially depends on the external conditions that were imposed to}. This simple idea has profound and unexpected implications. Let us illustrate the same one for the particular case of systems with negative heat capacities. Following the same arguments of Thirring \cite{Thirring}, let us consider the typical arrangement considered in statistical mechanics: a closed system composed of two finite systems A and B, which are put in thermal contact among them and isolated of any external influence.  As usual, let us assume that the total energy $U_{T}$ and entropy $S_{T}$ of the closed system are \emph{additive quantities}, $U_{T}=U_{A}+U_{A}$ and $S=S_{A}+S_{B}$. Maximization of entropy $S_{T}$ at constant energy $U_{T}$ demands the stationary condition:
\begin{equation}\label{stat.eq}
\frac{\partial S_{T}}{\partial U_{A}} =\frac{\partial S_{A}}{\partial U_{A}}-%
\frac{\partial S_{B}}{\partial U_{B}}=0\Rightarrow\frac{1}{T_{A}}=\frac {1}{%
T_{B}}\left(\equiv \frac{1}{T}\right),
\end{equation}
as well as the \emph{stability condition}:
\begin{equation}  \label{stab}
\frac{\partial^{2}S_{T}}{\partial U_{A}^{2}}=\frac{\partial^{2} S_{A}}{%
\partial U^{2}_{A}}+\frac{\partial^{2} S_{B}}{\partial U^{2}_{B}}
<0\Rightarrow\frac{C_{A}C_{B} }{C_{A}+C_{B}}>0,
\end{equation}
where $C_{A}$ and $C_{B}$ are heat capacities of these systems. Accordingly, systems A and B are found in thermal equilibrium if they exhibit the same temperature and their heat capacities satisfy one of the following stability conditions: (i) both systems exhibit positive heat capacities, or (ii) a system exhibits a negative heat capacity, e.g.: $C_{A}<0$, and the other a positive heat capacity $C_{B}>0$ that satisfies the following inequality \cite{Thirring}:
\begin{equation}
C_{B}<\left\vert C_{A}\right\vert .  \label{Thirring}
\end{equation}

According to the above result, a system with negative heat capacity is found in thermal equilibrium with other system (a bath) that exhibits a finite positive heat capacity. Conversely, thermal equilibrium is not possible if that system is put in thermal contact with a Gibbs thermostat (a heat bath that exhibits an infinite heat capacity). This type of equilibrium situation is the one associated with canonical or gran-canonical ensembles, which are widely employed to perform thermodynamical description of extensive systems of everyday applications. Precisely, these systems are under the influence of natural environment, whose heat capacity is regarded infinity for most of practical purposes. However, canonical or gran-canonical ensembles are unable to describes the existence of negative heat capacities. Since this anomaly is a relevant feature of astrophysical systems, these ensembles cannot provide an adequate treatment for them. In other words, thermo-statistical approaches of gravitational clustering that start from calculation of canonical partition function (\ref{ZZ}) are inadequate because of their consideration \emph{unable the access to fundamental information concerning to thermodynamics of astrophysical systems}.

\subsection{Violation of zeroth-law of thermodynamics}\label{violation}

\begin{figure}[tbp]
  \centering
  \includegraphics[width=2.5in]{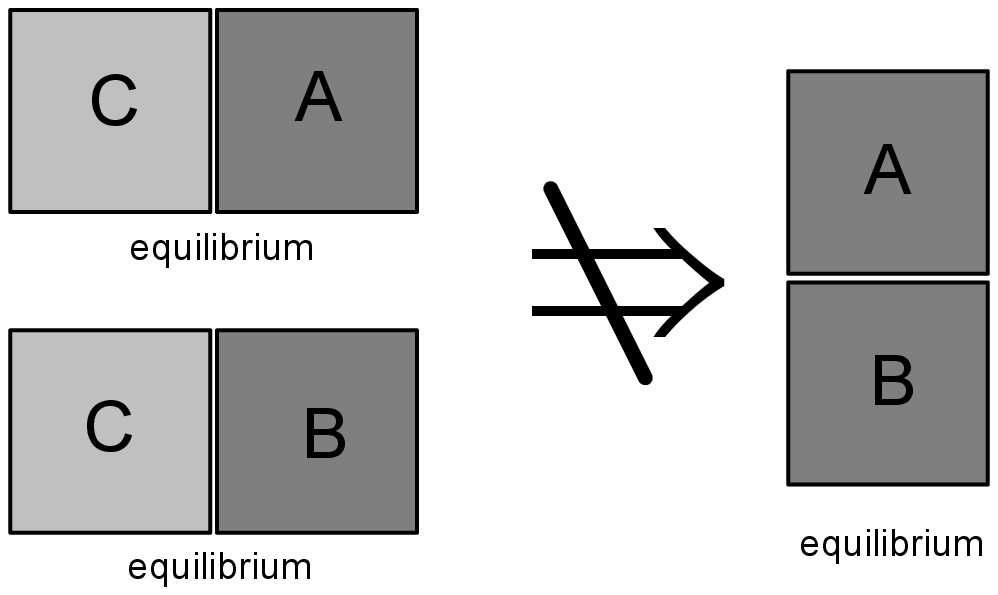}\\
  \caption{Violation of zeroth law of thermodynamics. Two systems A and B can be in equilibrium, by separate, with a third system C. However, this fact does not imply that the systems A and B will remain in equilibrium when they are put in thermal contact. There exist the following counterexample: the systems A and B have negative heat capacities $C_{A}<0$ and $C_{B}<0$, and the third system C a positive heat capacity $C_{C}<\min\left[\left\vert C_{A}\right\vert ,\left\vert C_{A}\right\vert \right]$.}\label{zeroth.eps}
\end{figure}

An unexpected consequence of the above analysis concerns to the called \emph{zeroth law of thermodynamics} \cite{Guggenheim}. This rule talks about transitivity of thermodynamic equilibrium: if two systems are both in thermal equilibrium with a third system then they are in thermal equilibrium with each other. Although this \emph{law} helps to define the experimental notion of temperature, its validity is restricted to conventional short-range extensive systems of everyday applications that exhibit nonnegative heat capacities (see scheme of figure \ref{zeroth.eps}). According to Thirring stability conditions, two identical systems that were initially prepared in the same macroscopic state, cannot remain in thermal equilibrium with each other if this state exhibits a negative heat capacity. This violation of zeroth law of thermodynamics was recently discussed by Ram\'{i}rez-Hern\'{a}ndez et al. \cite{Ramirez}. Curiously, these authors cited Thirring's paper \cite{Thirring}, but they didn't discuss the relevance of inequality (\ref{Thirring}) on explanation of this violation.

\subsection{Conciliating the notion of negative heat capacities within classical fluctuation theory}

Fluctuation relation (\ref{can.fr}) does not mean that the heat capacity is nonnegative. Its true meaning is that only those states with nonnegative heat capacities are \emph{stable} under the external conditions associated with canonical ensemble. According to Thirring stability criterium (\ref{Thirring}), open systems with negative heat capacities can be studied using a \emph{generalized statistical ensemble} that involves a thermal contact with a bath with finite positive heat capacity. This argument was considered by Curilef and I to generalize canonical fluctuation relation (\ref{can.fr}) as follows \cite{Vel.ETFR,Vel.Thirring}:
\begin{equation}\label{ETFR}
C=k\beta^{2}\left\langle\delta U^{2}\right\rangle+C\left\langle\delta \beta\delta U\right\rangle,
\end{equation}
which is now compatible with the existence of negative heat capacities $C<0$. As usual, $\delta x=x-\left\langle x\right\rangle$ denotes the thermal fluctuation of a given quantity $x$ with statistical expectation value $\left\langle x\right\rangle$. The presence of the term $\left\langle\delta \beta\delta U\right\rangle$ accounts for the feedback perturbation of the environmental (or the heat bath) inverse temperature $\beta$ as consequence of the energy interchange with the system. Conventional ensembles of equilibrium statistical mechanics (e.g.:, canonical or gran-canonical ensembles) disregard this type of influence\footnote{Besides negative heat capacities, the notion of \emph{temperature fluctuations} is also conflictive. Feshbach, Kittel and Mandelbrot protagonized in Physics Today a lively debated in turns to this concept in the past \cite{Feshbach,Kittel,Mandelbrot}. Kittel argumented that a system exhibits a \emph{definite temperature} when it is put in contact with a heat bath with infinite heat capacity. In other words, he regarded temperature as the control parameter $\beta=1/kT$ of canonical ensemble $p(U_{i}\vert \beta)=\exp(-\beta U_{i})/Z(\beta)$. Consequently, Kittel claimed that the notion of temperature fluctuations is an \emph{oxymoron}. Certainly, temperature parameter of canonical ensemble does not admit thermal fluctuations, in the same way that this ensemble does not describe negative heat capacities. However, microcanonical definition of temperature $1/T=\partial S/\partial U$ does not undergo these limitations.}. Applying this fluctuation relation to a system A that is put in thermal contact with a system B (environment), one obtains:
\begin{equation}  \label{fluct.deriv}
\frac{C_{A}C_{B}}{C_{A}+C_{B}}=k\beta^{2}\left\langle \delta
U_{A}^{2}\right\rangle .
\end{equation}
Stability condition (\ref{stab}) is now re-derived from positivity of right-hand side of equation (\ref{fluct.deriv}). Relation (\ref{fluct.deriv}) drops to canonical fluctuation relation (\ref{can.fr}) in the limit $C_{B}\rightarrow+\infty$, as well as the microcanonical result $\left\langle \delta U_{A}^{2}\right\rangle
\rightarrow0$ when $C_{B}\rightarrow0^{+}$. Accordingly, fluctuation relation (\ref{ETFR}) is associated with a family of generalized statistical ensembles that contains microcanonical and canonical ones as particular cases. Notice that these generalized ensembles can be implemented in a physical laboratory with a great accuracy.

Classical fluctuation theory that is discussed in most of textbooks of equilibrium statistical mechanics \cite{Reichl}, is a formulation with a restricted applicability. Most of theorems such as the canonical fluctuation relation (\ref{can.fr}) between the heat capacity $C$ and the energy $U$, and other known results as the fluctuation relation:
\begin{equation}
\chi_{T}=\beta\left\langle\delta M^{2}\right\rangle
\end{equation}
between the isothermal magnetic susceptibility $\chi_{T}$ and the total magnetization $M$, provide a wrong treatment to the existence of anomalous values in response functions. Derivation of these theorems invokes approximations like \emph{Gaussian approximation}, or alternatively, some special system-surrounding equilibrium situations as the ones associated with canonical or gran canonical ensembles, where the thermodynamic parameters of the surrounding (such its temperature $T$, pressure $p$, magnetic field $B$, chemical potential $\mu$, etc.) are regarded as \emph{unperturbed quantities} regardless the interaction with the system under consideration. However, a different perspective is revealed when one avoids the use of these arguments. For example, fluctuation relation (\ref{fluct.deriv}) was pioneering derived by Challa and Hetherington in the framework of Montecarlo simulations based on the Gaussian ensemble \cite{Challa,Hetherington}:
\begin{equation}
\omega_{G}\left( U\right) =\frac{1}{Z_{\lambda}\left( \beta_{s}\right) }\exp%
\left[ -\beta_{s} U-\lambda_{s}\left( U-U_{s}\right) ^{2}\right]
\label{GEns}
\end{equation}
with parameter $\lambda_{s}\geq 0$, which describes intermediate equilibrium situations between thermal contact with a Gibbs thermostat (canonical ensemble) when $\lambda_{s}\rightarrow 0^{+}$ and energy isolation (microcanonical ensemble) when $\lambda_{s}\rightarrow +\infty$. The form (\ref{ETFR}) is slightly more general, and yet, this is just a particular case of more general fluctuation theorems compatible with anomalous values in response functions. As already shown by Curilef and I, these theorems appear as first-order approximations of the following \textit{exact fluctuation theorem} \cite{Vel.GEN}:
\begin{equation}\label{exact.ft}
\left\langle\delta \eta_{i}\delta X^{j}\right\rangle=-k\delta^{i}_{i},
\end{equation}
which represents a fluctuation form of \textit{Le Chatelier-Brown principle} \cite{Reichl}: \emph{Any change in status quo prompts an opposing reaction in the responding system}. Here, $\eta_{i}=\partial S(X|\theta)/\partial X^{i}$ are the \emph{generalized restituting forces} derived from the entropy $S(X|\theta)$ of a closed system that appears in Einstein postulate \cite{Reichl}:
\begin{equation}\label{EP}
dp(X|\theta)=Ae^{S(X|\theta)/k}dX.
\end{equation}
Moreover, $k$ is the Boltzmann constant and $\theta$ certain control parameters for this closed system. Exact result (\ref{exact.ft}) is a direct consequence of the divergence theorem, which is surprising that the same one has not been derived before. Considering Cauchy-Schwartz inequality $\left|\left\langle \delta A\delta B\right\rangle\right|^{2}\leq \left\langle \delta A^{2}\right\rangle \left\langle \delta B^{2}\right\rangle$, this exact theorem leads to the following uncertainty-like inequality:
\begin{equation}\label{unc}
\Delta\eta_{i}\Delta X^{i}\geq k,
\end{equation}
with $\Delta x=\sqrt{\left\langle \delta x^{2}\right\rangle}$ being the statistical uncertainty of a random quantity $x$. Formally, this is a \emph{thermo-statistical analogous} of Heisenberg uncertainty relations of quantum mechanics \cite{Vel.Ann}, whose existence was postulated by Bohr in the early days of quantum mechanics \cite{Bohr}. By itself, this result suggests an interesting analogy between statistical mechanics and quantum mechanics \cite{Vel.Ann}, and therefore, an analogy between thermodynamics and classical mechanics.

\subsection{Conventional understanding of phase transitions and critical phenomena revisited}

A direct consequence of the above exact results is that the external conditions modify the thermodynamic stability and fluctuating behavior of a given system. by itself, this simple conclusion challenges the conventional understanding of phase transitions and critical phenomena \cite{Vel.Crit}. For example, the exact solution of 2D Ising model predicts a divergence of the heat capacity at the critical point \cite{Reichl}. According to canonical fluctuation relation (\ref{can.fr}), such a divergence implies the divergence of thermal fluctuations of energy. However, thermal fluctuations of energy described by equation (\ref{fluct.deriv}) now depend on the presence of bath with finite heat capacity $C_{B}$. This generalized fluctuation relation predicts that energy fluctuations remain finite regardless the divergence of system heat capacity $C_{A}\rightarrow+\infty$. This means that divergence of thermal fluctuations observed at the critical point, and other critical phenomena (e.g.: spontaneous breaking of symmetry, critical opalescence, long-range order, etc.), are not intrinsic properties for a given system because of their occurrence also depends on the external conditions.

\begin{figure}
\centering
\begin{minipage}{.5\textwidth}
  \centering
  \includegraphics[width=2.8in]{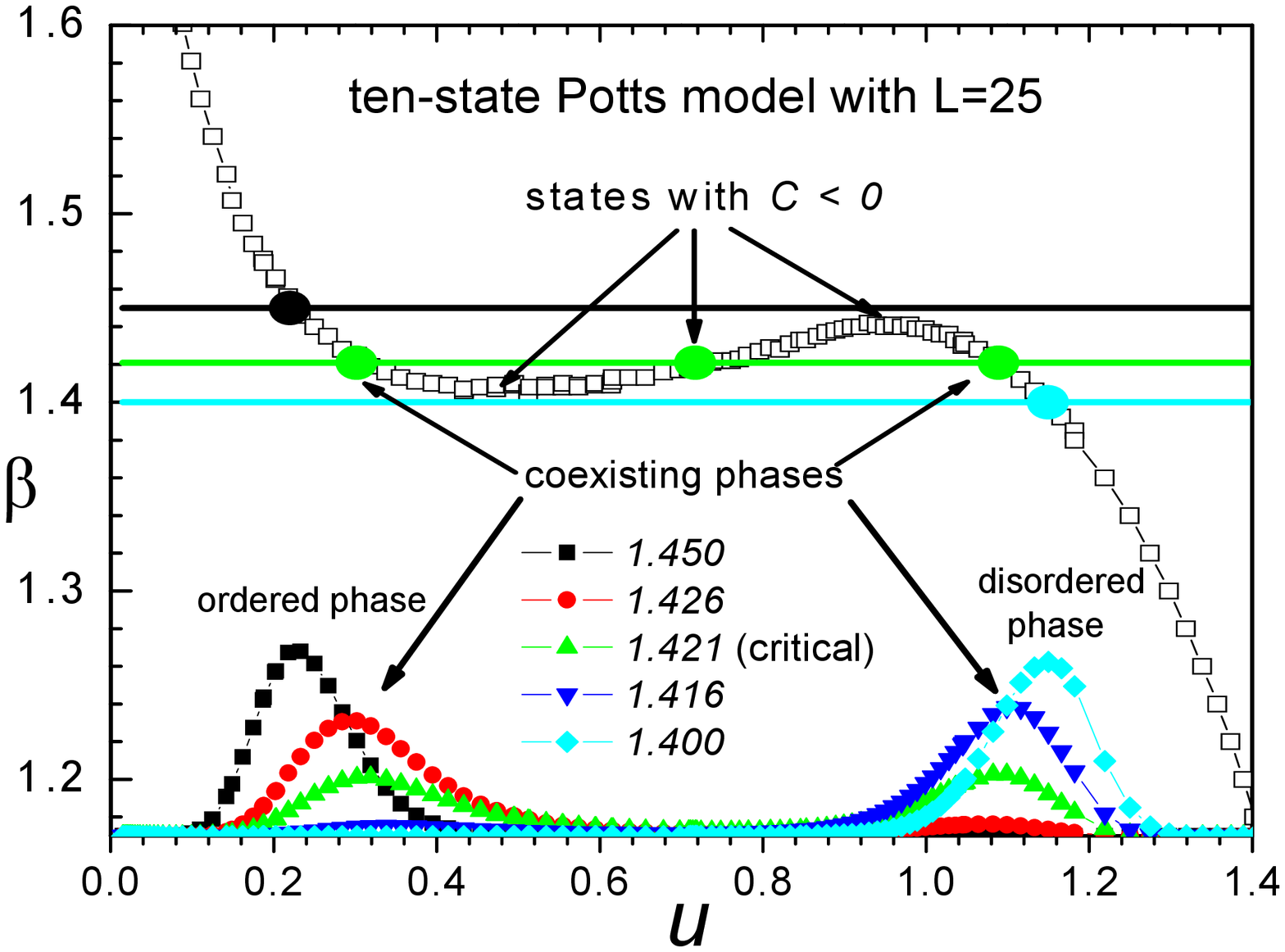}
\end{minipage}%
\begin{minipage}{.5\textwidth}
  \centering
  \includegraphics[width=2.8in]{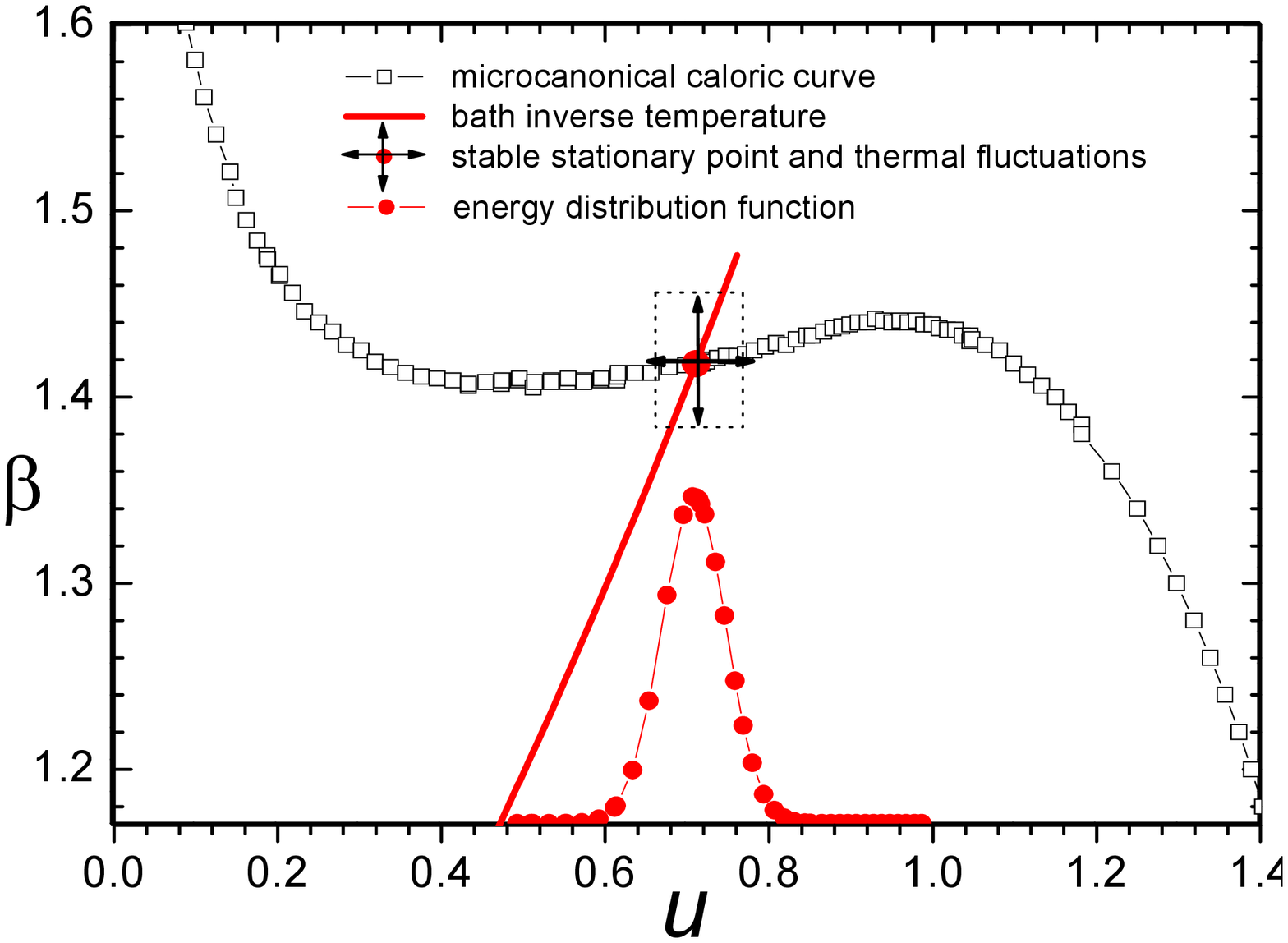}
\end{minipage}
\caption{Left panel: Behavior of energy distributions within canonical ensemble along the occurrence of phase coexistence phenomenon of the ten-state Potts model on the square lattice $25 \times 25$ with periodic boundary conditions ($u = U/N$ is the energy per site). This study clearly illustrates bimodal character of energy distribution functions when the inverse temperature parameter $\beta$ of canonical ensemble takes values around the critical value $\beta_{c}\simeq 1$.$421$ of temperature driven discontinuous PT. Notice that the branch of \emph{microcanonical caloric curve} $\beta(u)=\partial s(u)/\partial u$ (open squares) with states with negative heat capacities $C<0$ is poorly populated by using a bath with constant temperature, since these states are canonically unstable. The values of bath inverse temperature $\beta$ are represented here by horizontal lines. Intersection points of these horizontal lines with microcanonical inverse temperature correspond to the energies where energy distribution function exhibits its local maxima and minima. Right panel: The use of a bath with a finite heat capacity $C_{B}$ enables a direct study of the branch of microcanonical caloric curve (open squares) with negative heat capacities. Here, energy distribution exhibits a single Gaussian-peak that is located inside the region where microcanonical caloric curve exhibits negative heat capacities $C<0$. Both the bath inverse temperature $\beta_{B}$ (thick red line) and the system energy $U$ exhibit thermal fluctuations around to their equilibrium values, which are described by the energy-temperature fluctuation relation (\ref{ETFR}).}\label{potts.eps}
\end{figure}

Most of phase transitions of conventional extensive systems (e.g.: solid-liquid, liquid-gas, ferromagnetic-paramagnetic, etc.) are consequences of some type of thermodynamic instability associated with the presence of anomalous values in response functions. A simple example is shown in figure \ref{potts.eps}, specifically, the temperature-driven discontinuous phase transition of the ten-state Potts model on the square lattice $L\times L$ with periodic boundary conditions. As clearly shown in left panel of this figure, the branch of microcanonical caloric curve $\beta(U)=\partial S(U)/\partial U$ with negative heat capacities is non accessible within canonical ensemble. The presence of these states forces the \emph{bimodal character} of canonical energy distributions in the neighborhood of critical inverse temperature $\beta_{c}$ as well as the existence of a latent heat $q_{L}$ for the transition from the ordered phase towards the disordered one. These behaviors disappear once the heat bath of canonical ensemble is replaced by a heat bath with finite heat capacity $C_{B}$. Of course, the development of non-homogeneities, such as phase separations in fluids or formation of magnetic domains, will be also observed in this case, but all those effects associated with instability of negative heat capacities within canonical ensemble will disappear, such as bimodality of energy distributions, large thermal fluctuations, long-range statistical correlations, etc. Phase coexistence has nothing to do with the multimodal character of distributions. The formation of non-homogeneities actually occurs when there exist a unique energy peak, as the case shown in right panel of figure \ref{potts.eps}. The inverse temperature $\beta_{B}$ of the bath is not longer a good control variable for this type of situations, but the proper internal energy $U$ of the system. By adding a certain amount of energy to system-bath arrangement, the bath inverse temperature $\beta_{B}$ is self-adjusted to reach a thermal equilibrium. The mean values $\left\langle \beta_{B}\right\rangle$ and $\left\langle U\right\rangle$ are now fully able to describe the $S$-bend dependence of microcanonical curve of figure \ref{potts.eps}, while their fluctuation behavior enables a determination of states with negative heat capacities via fluctuation relation (\ref{ETFR}). This procedure have been employed with great success to design Monte Carlo methods much more efficient than others proposed in the literature \cite{Vel.MC}. The advantage is that several difficulties associated with occurrence of phase transitions can be avoided with a \emph{fine-tuning} of external conditions.

The present ideas have important implications in understanding the thermodynamics of astrophysical systems. As expected, the thermodynamic stability (or instability) of a given astrophysical system is not an intrinsic property of the same one since it depends on the external influence of other astrophysical systems. Surprisingly, such a dependence on the external conditions is not restricted to its fluctuating behavior and thermodynamic stability: \emph{every thermodynamic behavior} of the astrophysical system is considerably affected by any external perturbation because of the \emph{long-range character} of gravitation. This peculiarity is one of the reasons that explains why thermodynamics of astrophysical systems is so radically different from that of extensive systems of everyday practice.

\section{Thermodynamics of astrophysical systems through some simple models}

\subsection{Eddington arguments using virial theorem}\label{Eddington}

To my knowledge, the pioneering evidence about the relevance of negative heat capacities in astrophysics was provided by Eddington \cite{Eddington}. He  considered the virial theorem:
\begin{equation}\label{virial}
3pV=2\left\langle K \right\rangle+n\left\langle W\right\rangle
\end{equation}
for a 3D system of non-relativistic point particles whose interacting forces exhibit a radial dependence $1/r^{n+1}$. For open astrophysical systems $n=1$ and its external pressure $p=0$. Therefore, the expectation values of the total kinetic and potential energies are related as $\left\langle K\right\rangle=-\left\langle W\right\rangle/2$, while the total energy $\left\langle U\right\rangle=\left\langle K\right\rangle+\left\langle W\right\rangle=-\left\langle K\right\rangle$. Assuming the equipartition relation between kinetic energy and temperature $\left\langle K\right\rangle=3NkT/2$, one obtains the following estimation for heat capacity:
\begin{equation}
\left\langle U\right\rangle=-3NkT/2\rightarrow C=d\left\langle U\right\rangle/dT=-3Nk/2<0.
\end{equation}
Although the virial theorem (\ref{virial}) provides a simplest argument that the heat capacity of astrophysical systems could be negative, the general applicability of this analysis cannot be over-estimated. For example, the heat capacity can be positive when an astrophysical system is confined under an external potential well $w(\mathbf{r})$. On the other hand, the equipartition relation $\left\langle K\right\rangle=3NkT/2$ does not hold if the astrophysical system suffers from the evaporation of its constituents. Let us review some simple thermo-statistical astrophysical models that exemplify the present limitations of Eddington analysis.

\begin{figure}[tbp]
\begin{flushright}
\includegraphics[width=4.0in]{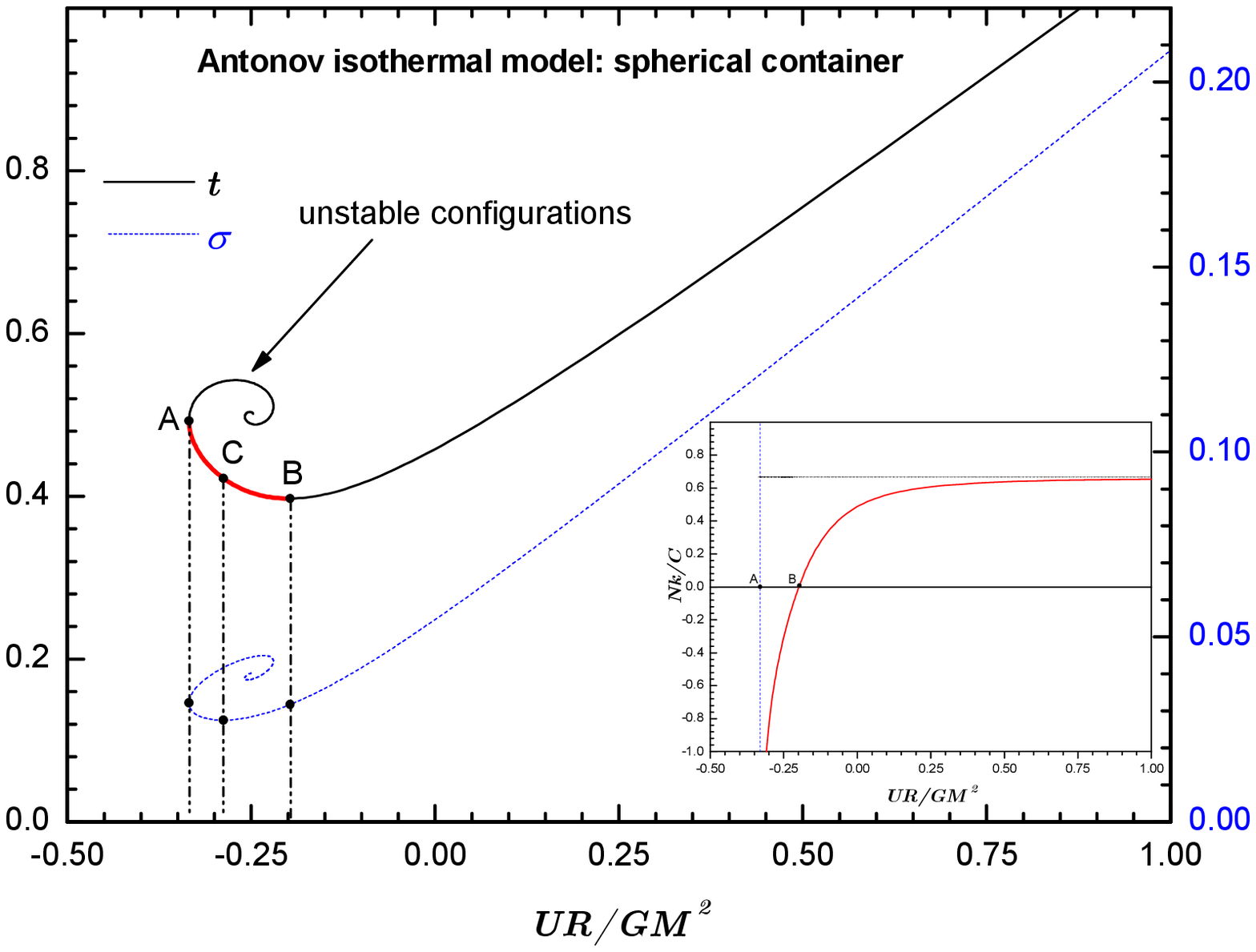}
\end{flushright}
\caption{Main panel: Microcanonical temperature and pressure dependencies associated with Antonov isothermal model with spherical container using dimensionless variables $u=UR/GM^{2}$, $t=kTR/GMm$ and $\sigma=pR^{4}/GM^{2}$. The notable points of caloric curve $u_{A}=-0$.$355$ and $u_{B}=-0$.$197$ correspond to critical points of gravothermal and isothermal collapses. The branch $u_{A}<u<u_{B}$ (red line) are configurations with negative heat capacities. The microcanonical pressure exhibits a third notable point $u_{C}=-0$.$288$ where the dimensionless pressure variable $\sigma$ over exhibits its minimal value $\sigma_{C}=0$.$027$. Inset panel: energy dependence of the inverse of the specific heat $c=C/Nk$. Notice that the asymptotic limit $c\rightarrow 3/2$ when $u\rightarrow +\infty$ (ideal gas behavior), and its vanishing when $u\rightarrow u^{+}_{A}$.}\label{Antonov.eps}
\end{figure}

\subsection{Antonov isothermal model}

The simplest model that considers the presence of a potential well is the Antonov isothermal model \cite{Antonov}. Specifically, this model considers a self-gravitating gas of identical non-relativistic classical point particles:
\begin{equation}\label{sgg}
H_{N}(\mathbf{r},\mathbf{p})=\sum_{i}\frac{1}{2m}\mathbf{p}^{2}_{i}+w_{c}(\mathbf{r}_{i})-\sum_{i<j}\frac{Gm^{2}}{\left|\mathbf{r}_{i}-\mathbf{r}_{j}\right|}
\end{equation}
that is enclosed inside a \emph{container with a impenetrable wall}:
\begin{equation}\label{impenetrable}
w_{c}(\mathbf{r})=\left\{
\begin{array}{cc}
0 & \mbox{if } \mathbf{r}\in \Omega\subset \mathds{R}^{3}\\
+\infty & \mbox{otherwise.} \\
\end{array}
\right.
\end{equation}
Since this external field exhibits an infinite potential barrier, it fully suppresses the evaporation of system constituents. Therefore, the equipartition relation $\left\langle K\right\rangle=3NkT/2$ holds as well as virial theorem (\ref{virial}) with non-vanishing external pressure $p$. For energies sufficiently large, the average potential energy $\left\langle W\right\rangle$ can be neglected in comparison with the average total kinetic energy $\left\langle K\right\rangle$, so that, this system behaves in this limit as an ideal gas with positive heat capacity $C=3Nk/2$ that fills uniformly the volume of container. For energies sufficiently low, the absolute value of the total gravitational potential and the kinetic energies become comparable, but the external pressure $p$ falls significantly. Accordingly, one should observe configurations with negative heat capacities for low energies. The simplest container geometry with boundary $\partial\Omega$ is the spherical shape with radius $R$, whose microcanonical dependencies are shown in figure \ref{Antonov.eps}. It was introduced here the dimensionless variables for energy $u=UR/GM^{2}$, temperature $t=kTR/GMm$ and pressure $\sigma=pR^{4}/GM^{2}$. Besides the qualitative behaviors previously commented, one observes the existence of three notable energies: $u_{A}=-0$.$335$, $u_{B}=-0.197$ and $u_{C}=-0$.$288$.

According to these dependencies, equilibrium configurations are possible for energies $u>u_{A}$ only. Therefore, a system with energy $u\leq u_{A}$ becomes unstable under its own gravitation field and undergoes a \emph{gravitational collapse}. For energies near the critical energy $u_{A}$, the system exhibits a branch with negative heat capacities up to the point $u_{B}$. Additionally, equilibrium configurations are possible for temperatures $t\geq t_{A}$ only, where $t_{A}=0$.$397$ is the temperature that corresponds to the second notable energy $u_{B}$. If the astrophysical system is put in \emph{thermal contact} with an heat bath of constant temperature $T$ (canonical ensemble), the same one will undergo a gravitational collapse if the bath temperature $T<T_{B}=t_{B}GMm/Rk$. This second instability process is referred to as \emph{isothermal collapse}, so that, the energy $u_{A}$ is the critical point of isothermal collapse. Finally, equilibrium configurations are only possible for pressures $\sigma\geq \sigma_{C}$, where $\sigma_{C}=0$.$027$ is the pressure that corresponds to the third notable energy $u_{C}$. If the radius $R$ of container is allowed to change and the system is surrounded by an environment with constant pressure $p$ (isobaric ensemble), the system will undergo a gravitational collapse when the environmental pressure $p<p_{C}=\sigma_{C}GM^{2}/R^{4}$. As expected, this third gravitational instability could be referred to as an \emph{isobaric collapse}. The practical realization of isothermal and isobaric collapses are quite unrealistic in the astrophysical scenario. As already discussed, thermal contact associated with canonical ensemble involves short-range interactions, but astrophysical systems interact among them through gravitation, which is a long-range force. Although the presence of a container is a simplifying assumption to study astrophysical systems, it is clearly nonphysical for real situations. The only gravitational instability that plays a relevant role in real applications is the gravothermal collapse. This process drives the formation of astrophysical structures in every scale, ranging from the formation of planetesimals up to the large-scale galaxies distribution in the Universe.

\subsection{Incidence of evaporation: King model}\label{King.section}

The incidence of evaporation represents an unavoidable process in the framework of real astrophysical systems. A simple model that accounts for the incidence of evaporation was proposed by King \cite{King}, which provides a suitable description about structure and evolution of star clusters in astrophysics. Its key assumption is the relevance of quasi-stationary one-body distribution:
\begin{equation}\label{MK.pdf}
f_{qs}(\mathbf{r},\mathbf{p}|\beta,\varepsilon_{c})=A\left\{e^{\beta\left[\varepsilon_{c}-\varepsilon\left(\mathbf{r},
\mathbf{p}\right)\right]}-1\right\}\Theta\left[\varepsilon_{c}-\varepsilon\left(\mathbf{r},
\mathbf{p}\right)\right],
\end{equation}
which was introduced by Michie \cite{Michie}. Formally, this is deformed Maxwell-Boltzmann profile that is truncated at the escape energy $\varepsilon_{c}$. Here,  $\varepsilon\left(\mathbf{r},\mathbf{p}\right)=\mathbf{p}^{2}/2m+m\varphi (\mathbf{r})$ is the mechanical energy of individual stars, while $\Theta(x)$ is the Heaviside step function. The energy parameter $\varepsilon_{c}<0$ represents the energy threshold of stars to escape from the gravitational field of globular cluster and fall into gravitational influence of a nearby galaxy.

\begin{figure}
\centering
\begin{minipage}{.5\textwidth}
  \centering
  \includegraphics[width=2.8in]{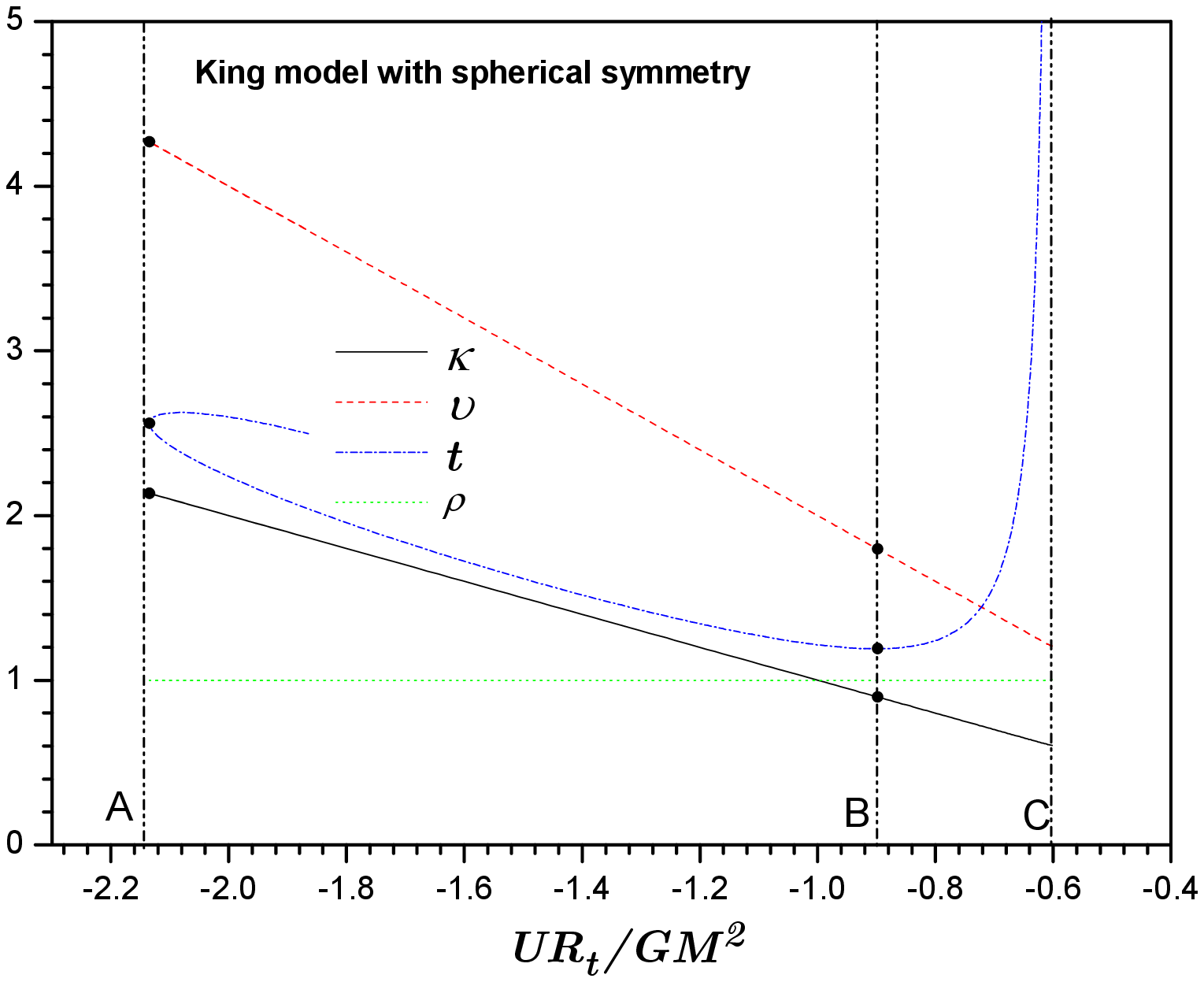}
\end{minipage}%
\begin{minipage}{.5\textwidth}
  \centering
  \includegraphics[width=2.8in]{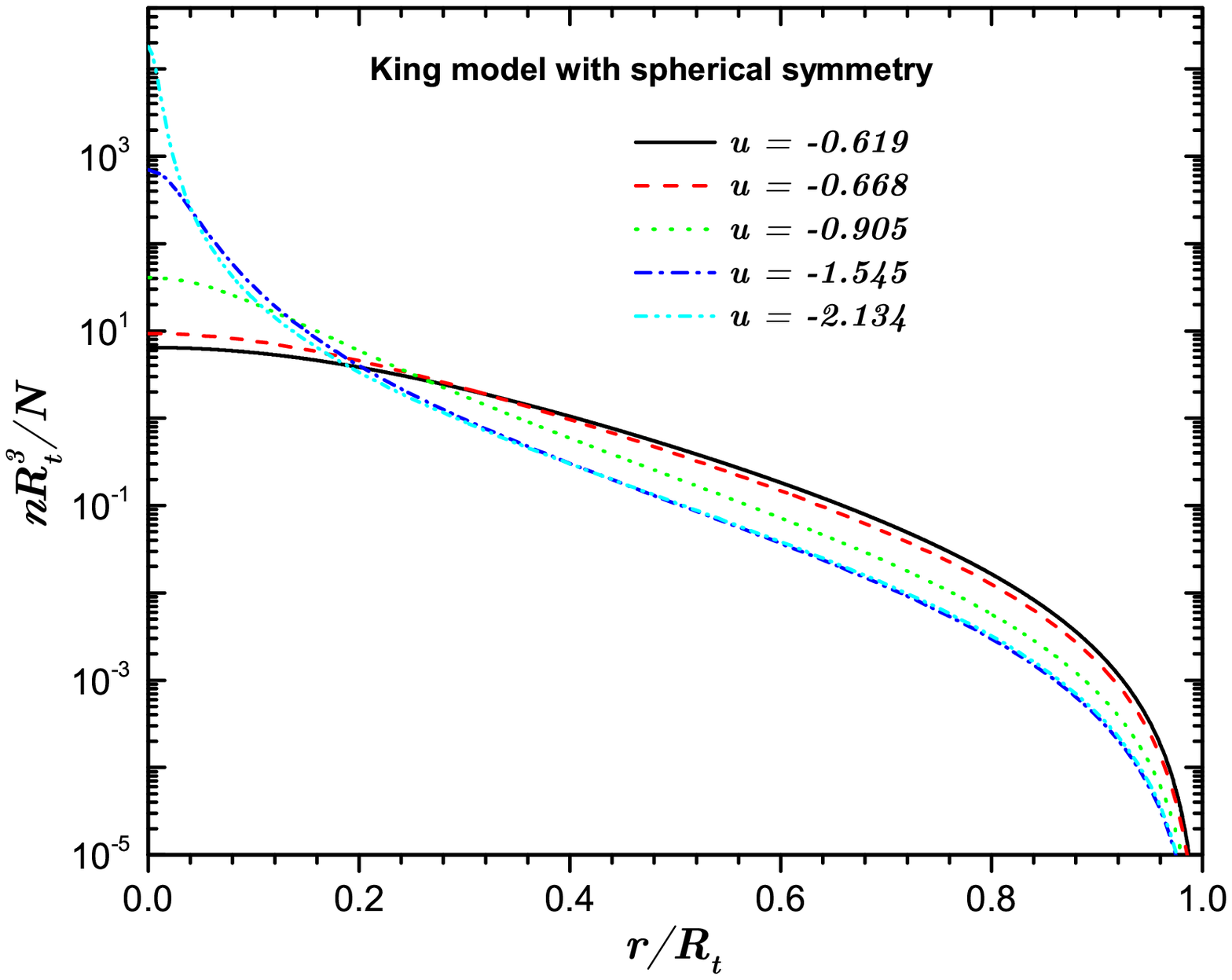}
\end{minipage}
\caption{Left panel: Dependence of some quantities \emph{versus} dimensionless energy $u=UR_{t}/GM^{2}$ associated with King model: dimensionless kinetic energy $\kappa=\left\langle K\right\rangle R_{t}/GM^{2}$, the negative value of the dimensionless potential energy $v=-\left\langle W\right\rangle R_{t}/GM^{2} $, the temperature-like parameter $t=kTR_{t}/GMm$ of one-particle quasi-stationary distribution (\ref{MK.pdf}); and the virial ratio $\rho=-2\left\langle K\right\rangle/\left\langle W\right\rangle$. Vertical dash-dot lines indicate the position of three notable points $(u_{A},u_{B},u_{C})$ of this model. Right panel: Distribution profiles associated with King model for different values of energy. Notice that the decreasing of the energy variable $u=UR_{t}/GM^{2}$ leads to distribution profiles with more dense cores and diluted haloes.}\label{King.eps}
\end{figure}

The one-body distribution (\ref{MK.pdf}) leads to the following particles density $n(\mathbf{r})$:
\begin{equation}\label{MK.density}
n(\mathbf{r})=\int f_{qs}(\mathbf{r},\mathbf{p}|\beta,\varepsilon_{c}) \frac{d^{3}\mathbf{p}}{\left(2\pi\hbar\right)^{3}}=C\exp{\left(\Phi\right)}F\left(\sqrt{\Phi}\right).
\end{equation}
Here, $C=A\left(m/2\pi\hbar^{2}\beta\right)^{3/2}$ is the normalization constant, $F(x)$ is the auxiliary function:
\begin{equation}\label{F(x)}
F(x)=\mathrm{erf}(x)-\frac{2}{\sqrt{\pi}}\left(x+\frac{2}{3}x^{3}\right)e^{-x^{2}},
\end{equation}
while $\mathrm{erf}(x)$ is the error function:
\begin{equation}\label{error.func}
\mathrm{erf}(x)=\frac{2}{\sqrt{\pi}}\int_{0}^{x}e^{-t^{2}}dt.
\end{equation}
Finally, $\Phi=\beta\left[\varepsilon_{c}-m\varphi(\mathbf{r})\right]$ denotes the dimensionless potential. For an spherical symmetric distribution, expression (\ref{MK.density}) predicts the vanishing of stars distribution $n(\mathbf{r})$ outside the region:
\begin{equation}
\left|\mathbf{r}\right|\leq R_{t}=-GMm/\varepsilon_{c},
\end{equation}
with $R_{t}$ being the tidal radius \cite{Binney}. Here, $M$ is the total mass of the system, while $m$ is the mass of its constituting stars. The tidal radius $R_{t}$ appears as an independent parameter of King model, but it can be estimated by analyzing the gravitational field of the nearby galaxy. For example, the estimation of equation (\ref{tidal.quad}) is derived in the framework of quadrupole approximation (\ref{quadrupole}) by assuming that a globular cluster of mass $M_{A}$ moves along a circular orbit of average distance $D$ around a galaxy of mass $M_{B}$.

According to dependencies shown in left panel of figure \ref{King.eps}, King model satisfies the virial ratio $\rho=-2\left\langle K\right\rangle/\left\langle W\right\rangle\equiv 1$, which implies a decreasing of kinetic energy when the total energy $U$ is increased. However, the mean value of total kinetic energy $\left\langle K\right\rangle$ does not follow a linear relation with the quantity $T=1/k\beta$, which could be regarded as the \emph{temperature parameter} of quasi-stationary one-body distribution (\ref{MK.pdf}). Once again, one observes a branch of caloric curve with negative heat capacities $C<0$ between the point of gravothermal collapse $u_{A}\simeq-2$.$134$ and the point of isothermal collapse $u_{B}=-0.898$. A new feature is the upper bound for the total energy at $u_{C}=-0.601$, which corresponds to the point of \emph{evaporation disruption}. Accordingly, a macroscopic configuration with energy $u\geq u_{C}$ is unstable under evaporation. Such an astrophysical system will undergo a sudden evaporation in order to release its excess of energy and gain stability. It is remarkable that the energy of gravothermal collapse $u_{A}$ predicted by King model is appreciably lower than the one corresponding to Antonov isothermal model. This behavior is related to the vanishing of particles distributions at the tidal radius $R_{t}$. It is shown in right panel of figure \ref{King.eps} that profiles with more diluted haloes also exhibit more dense cores. Consequently, the absolute values of the total potential energy and its associated kinetic energies and temperatures are greater than the ones corresponding to the isothermal model. Several aspects of thermodynamics of King models and related questions have been recently analyzed in the literature by other authors. Interested readers can see references \cite{CasettiMK,ChavanisMK} for further details.

\subsection{Generalization: Truncated $\gamma$-exponential models}

\begin{figure}
\centering
\begin{minipage}{.5\textwidth}
  \centering
  \includegraphics[width=2.8in]{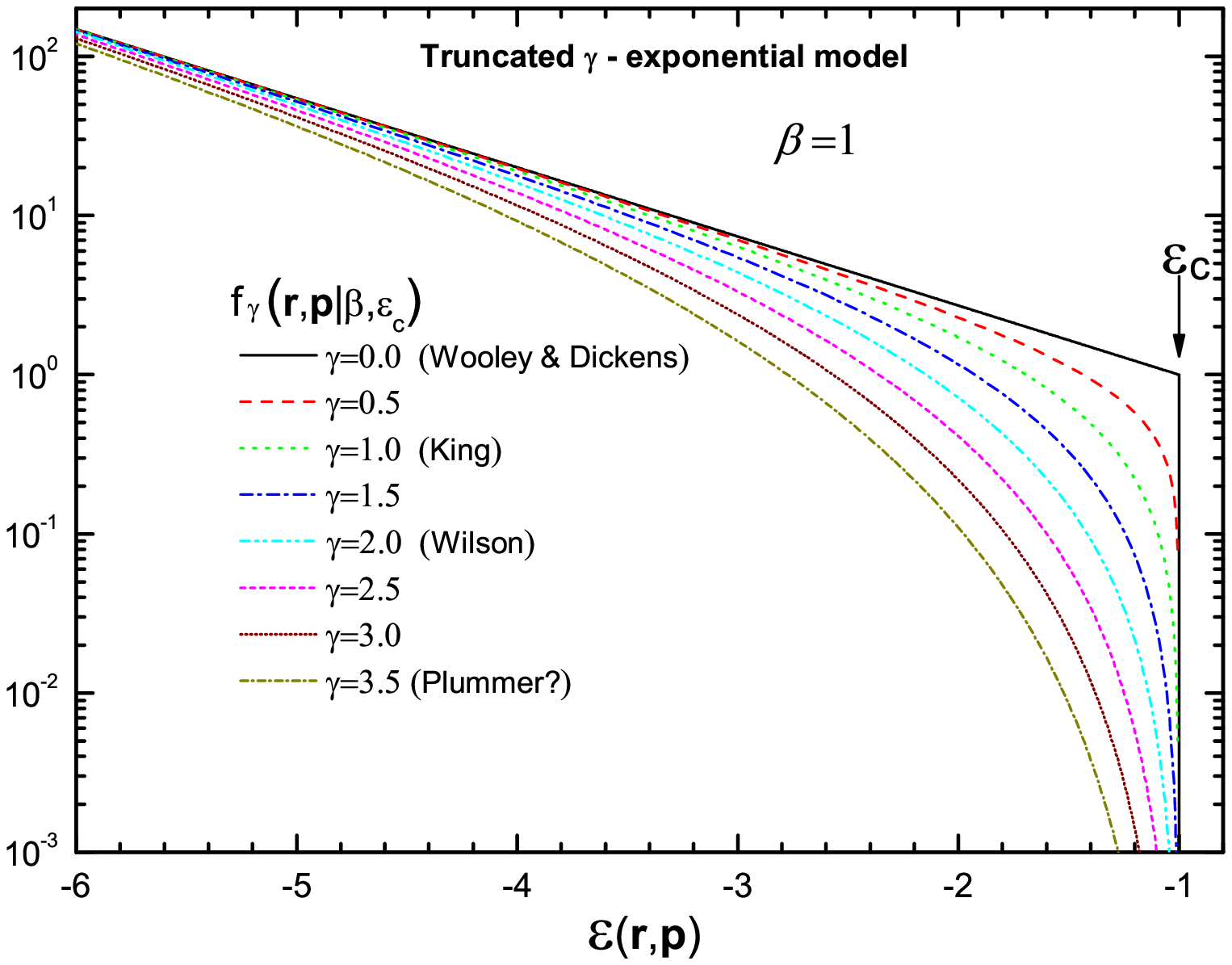}
\end{minipage}%
\begin{minipage}{.5\textwidth}
  \centering
  \includegraphics[width=2.8in]{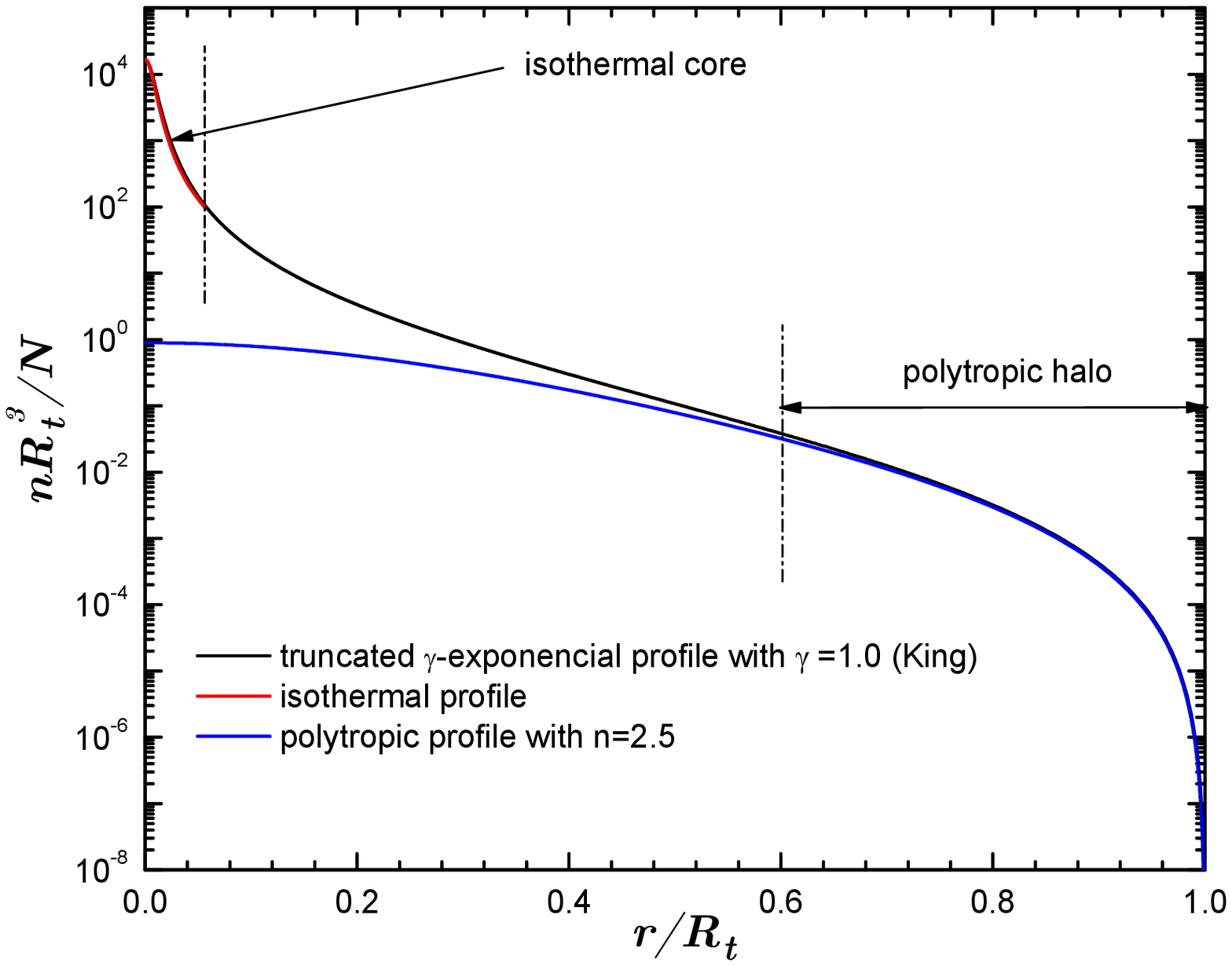}
\end{minipage}
\caption{Left panel: One-particle distributions \emph{versus} mechanical energy $\varepsilon(\mathbf{q},\mathbf{p})$ corresponding to truncated $\gamma$-exponential model (\ref{df.model}) for some values of deformation parameter $\gamma$. It was employed the prefixed values $\varepsilon_{c}=-1$ and $\beta=1$ for the escape energy and the inverse temperature, respectively. Right panel: Truncated $\gamma$-exponential profile with $\gamma = 1$ (King model) corresponding to the point of gravothermal collapse and its comparison with isothermal and polytropic profiles using log-linear scales.}\label{gamma.distribution.eps}
\end{figure}

Besides King model, the incidence of evaporation is also considered by other simple models, such as the truncated isothermal model of Wooley and Dickens \cite{Wooley}, and Wilson model \cite{Wilson}. Quasi-stationary one-body distributions of all these models can be expressed into a unifying form as follows:
\begin{equation}\label{df.model}
f_{qs}(\mathbf{r},\mathbf{p}|\beta,\varepsilon_{c})=A E_{t}(x),
\end{equation}
where $x=\beta\left[\varepsilon_{c}-\varepsilon(\mathbf{r},\mathbf{p})\right]$, and $E_{t}(x)$ is a certain \emph{truncated exponential function} $E_{t}(x)$ that vanishes for $x<0$. As already recently discussed by Gomez-Leyton and I \cite{Vel.Trunc}, all these models belong to a parametric family of models that employs the \emph{$\gamma$-exponential function}:
\begin{equation}\label{integral.gamma}
E_{t}\left(x\right)\rightarrow E\left(x;\gamma\right)=
\sum_{k=0}^{\infty}\frac{x^{\gamma+k}}{\Gamma\left(
\gamma+1+k\right)}.\\
\end{equation}
This function is obtained from \emph{fractional integration} of ordinary exponential function \cite{Miller}
\begin{equation}\label{integral.gamma}
E\left(x;\gamma\right)=\frac{1}{\Gamma(\gamma)}\int^{x}_{0}\eta^{\gamma-1}\exp(x-\eta)d\eta,
\end{equation}
where $\Gamma(x)$ is the known Euler Gamma function. It converges towards ordinary exponential function for $x>0$ sufficiently large, but it present a power-law dependence $x^{\gamma}$ for $x\rightarrow+0$. The behavior of these distributions for different values of deformation parameter $\gamma$ are shown in left panel of figure \ref{gamma.distribution.eps}. This family of phenomenological models provides us a more complete picture about the incidence of evaporation on the thermodynamics of astrophysical systems. It contains Wooley and Dicken model for $\gamma=0$, King model for $\gamma=1$, Wilson model for $\gamma=2$, while Plummer model \cite{Plummer} seems to be a singular asymptotic case when $\gamma\rightarrow7/2$ and $\varepsilon_{c}\rightarrow0^{-}$.

The relevance of deformation parameter $\gamma$ becomes evident once obtained the particles density $n(\mathbf{r})$:
\begin{equation}\label{den}
n(\mathbf{r})=\int f_{qs}(\mathbf{r},\mathbf{p}|\beta,\varepsilon_{c}) \frac{d^{3}\mathbf{p}}{\left(2\pi\hbar\right)^{3}}=C E\left(\Phi;\gamma+\frac{3}{2}\right),
\end{equation}
where, $C=A\left(m/2\pi\hbar^{2}\beta\right)^{3/2}$ is the normalization constant. Notice that particles density is also expressed by a truncated exponential function. This profile describes a power-law dependence for $\Phi$ small with exponent $n=\gamma+3/2$, and an exponential dependence for $\Phi$ large:
\begin{equation}\label{asymp.density}
n\left(  \Phi;\gamma\right)  =\left\{
\begin{array}
[c]{cc}%
\propto\Phi^{n}, & \Phi\ll1,\\
\propto\exp\left(  \Phi\right)  , & \Phi\gg1.
\end{array}
\right.
\end{equation}
Therefore, particles distribution (\ref{den}) unifies the known isothermal and polytropic profiles with $n=\gamma+3/2$. The deformation parameter $\gamma$ determines the \emph{polytropic structure} of the haloes, being this feature a consequence of evaporation. Some of these profiles are shown in right panel of figure \ref{gamma.distribution.eps}, which has been fitted using an isothermal profile for the core and a polytropic profile for the halo. Since the deformation parameter $\gamma$ is nonnegative and the polytropic profiles are unstable for $n\geq 5$ \cite{Chandrasekhar}, the admissible values of the polytropic exponent $n$ are restricted to the interval $3/2\leq n<5$, which correspond to the interval $0\leq \gamma < 7/2$ for the deformation parameter $\gamma$. Distribution profiles are obtained from numerical integration of the following nonlinear Poisson problem:
\begin{equation}\label{p1}
\frac{1}{\xi^{2}}\frac{d}{d\xi}\left[\xi^{2}\frac{d}{d\xi}\Phi(\xi)\right]=-4
\pi E\left[  \Phi(\xi);\gamma+\frac{3}{2}\right]
\end{equation}
in terms of the dimensionless radius $\xi=\sqrt{\kappa}r/R_{t}$ by demanding the following conditions at the origin:
\begin{equation}
\Phi\left(  0\right)  =\Phi_{0}\mbox{ and }\frac{d}{d\xi}\Phi\left(  0\right) =0.
\end{equation}
The auxiliary parameter $\kappa=\xi^{2}_{c}$ and the inverse temperature $\eta=\beta GMm/R_{t}$ are determined from vanishing of dimensionless potential:
\begin{equation}\label{Boundary}
\Phi\left(  \xi_{c}\right)  =0\mbox{ and }\xi_{c}\frac{d}{d\xi}\Phi\left( \xi_{c}\right)=-\eta.
\end{equation}

\begin{figure}[tbp]
\begin{flushright}
\includegraphics[width=6.0in]{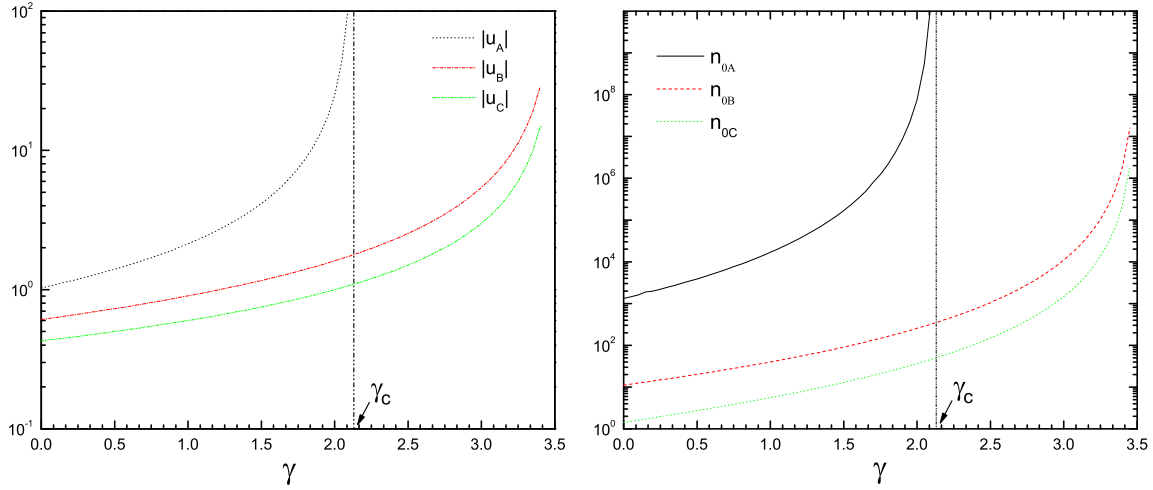}
\end{flushright}
 \caption{Behavior of dimensionless energy and central density at the three notable energies. Notice that the divergence of $|u_{A}|$ and $n_{0A}$ at the notable value $\gamma_{c}\simeq 2$.$13$, and the divergence of the other quantities when $\gamma\rightarrow7/2$.}\label{detallado.eps}
\end{figure}

\begin{figure}[tbp]
\begin{flushright}
\includegraphics[width=4.0in]{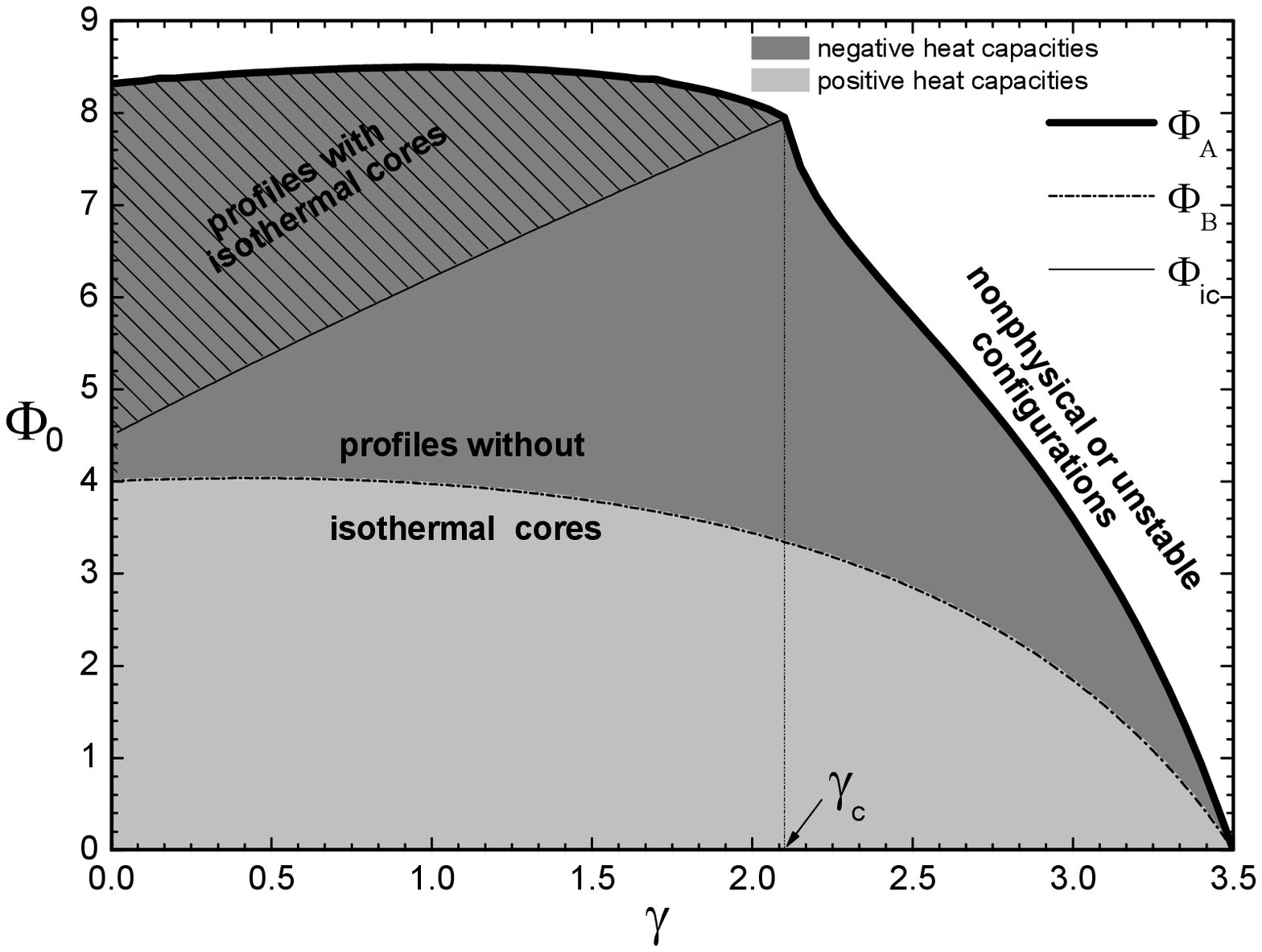}
\end{flushright}
\caption{Phase diagram of truncated $\gamma$-exponential models in
the plane of integration parameters $\gamma-\Phi_{0}$. After \cite{Vel.Trunc}.}\label{PhiConverge.eps}
\end{figure}

It is shown in figure \ref{detallado.eps} the influence of deformation parameter $\gamma$ on the dimensionless energy $u=UR_{t}/GM^{2}$ and central density $n_{0}=n\left(0\right)R^{3}_{t}/N$ for notable energies of microcanonical caloric curve. Notice that it has been employed here the same notation convention for notable points of caloric curve shown in figure \ref{King.eps}. With a greater influence of evaporation, stable configurations exhibit distribution profiles with cores more dense and hotter, and simultaneously, more diluted haloes. It is remarkable that the existence of a notable value of deformation parameter $\gamma_{c}\simeq2$.$13$, where one observes the divergence of energy and density corresponding to the gravothermal collapse, $u_{A}=+\infty$ and $n_{0A}=+\infty$ when $\gamma_{c}\leq \gamma<7/2$. Since the system needs to release an infinitely amount of energy to reach this ground state, the dramatic character of gravothermal collapse has been considerably reduced in comparison to the cases $0\leq\gamma<\gamma_{c}$. The explanation of this behavior can be understood by analyzing the figure \ref{PhiConverge.eps}, which represents the \emph{phase diagram} in the plane of integration parameters $\gamma-\Phi_{0}$ of differential equation (\ref{p1}). The central values of dimensionless potential $\Phi_{0}$ provides a measure about the isothermal character of the core. According to asymptotic behavior (\ref{asymp.density}), larger values of this parameter guarantee the existence of isothermal cores, whose maximum value $\Phi_{A}$ corresponds to the point of gravothermal collapse. The same one exhibits a very weak dependence on the deformation parameter $\gamma$ for the cases $0\leq\gamma<\gamma_{c}$, and suddenly, it starts to decreases for the cases $\gamma_{c}\leq \gamma<7/2$. The inner regions of distribution profiles with $\gamma_{c}\leq \gamma<7/2$ are so cuspid that they cannot be adjusted with the isothermal profile. Isothermal cores are only possible for low energies and values of deformation parameter $\gamma$ below the notable value $\gamma_{c}$, $0\leq\gamma<\gamma_{c}$. Finally, one can observe a progressive divergence of dimensionless energy $\left|u\right|$ and central density of the other notable points when $\gamma\rightarrow 7/2$, which is directly related to the instability of polytropic profiles with $n\geq 5$.

\subsection{Isolated astrophysical system: Plummer model}

The polytropic model with $n=5$ exhibit analytical solutions for both particles density:
\begin{equation}\label{Plumer.prof}
n(\mathbf{r})=\left(\frac{3N}{4\pi a^{3}}\right)\left(1+\frac{r^{2}}{a^{2}}\right)^{-5/2}
\end{equation}
and the gravitational potential:
\begin{equation}
\varphi(\mathbf{r})=-\frac{GM}{\sqrt{r^{2}+a^{2}}}
\end{equation}
with $M=Nm$. Although this distribution is infinitely extended in the space, it exhibits a finite total mass $M$ due to the fast power-law decreasing $1/r^{5}$ when $r\rightarrow+\infty$. Originally, this model was introduced by Plummer \cite{Plummer} to fit distribution profiles of star clusters, but nowadays is employed in $N$-body simulations as a toy model to assign initial conditions \cite{Aarseth}. This model can be associated with the following quasi-stationary one-body distribution:
\begin{equation}\label{Plumer.model}
f_{P}(\mathbf{r},\mathbf{p}|\beta)=A\left[-\beta\varepsilon(\mathbf{r},\mathbf{p})\right]^{7/2}.
\end{equation}
This distribution seems to be an asymptotic case of truncated $\gamma$-exponential model (\ref{df.model}) with $\gamma\rightarrow7/2^{-}$ and threshold energy $\varepsilon_{s}\rightarrow0^{-}$ (or tidal radius $R_{t}\rightarrow+\infty$). Calculating the relation between the characteristic length $a$ and the total energy:
\begin{equation}
U=\int \left[\frac{1}{2m}\mathbf{p}^{2}+\frac{1}{2}m\varphi(\mathbf{r})\right]
f_{P}(\mathbf{r},\mathbf{p}|\beta)
\frac{d^{3}\mathbf{r}d^{3}\mathbf{p}}{\left(2\pi\hbar\right)^{3}},
\end{equation}
one can show that the normalization constant $A$ can be expressed as follows:
\begin{equation}\label{AA.Plummer}
A=\frac{2^{8}3}{7}\left(\frac{\hbar^{2}}{2Gm^{3}aN^{1/3}}\right)^{3/2}\left(\frac{\pi}{32}\right)^{7/2},
\end{equation}
where the characteristic length $a$ and the inverse temperature $\beta$ are related to the total energy $U$ as:
\begin{equation}\label{Plummer.length}
a=\frac{3\pi}{64}\frac{GM^{2}}{\left(-U\right)}\mbox{ and }\beta=\frac{32}{\pi}\frac{a}{GMm}\equiv\frac{3N}{2\left(-U\right)}.
\end{equation}
This model satisfies virial theorem (\ref{virial}) with vanishing external pressure $p=0$. Therefore, it describes an open astrophysical system with no external gravitational influence. All its properties are determined by its internal energy $U$ and number of particles $N$. Notice that this model does not describe the occurrence the gravitational collapse and the evaporation disruption.

\subsection{Quasi-ergodic model}

Among the astrophysical models reviewed here, Antonov isothermal model is the only based on equilibrium statistical mechanics. The other models represent  phenomenological attempts to include evaporation effects. Even the expression (\ref{MK.pdf}) was a suitable approximation introduced by Michie \cite{Michie}. Let us finally review a simple astrophysical model that also includes evaporation effects, but this time from a purely statistical mechanics perspective: the \emph{quasi-ergodic model} \cite{Vel.QEM}. Its introduction invokes the existence of a certain quasi-ergodicity of microscopic dynamics, that is, the system evolution is able to perform a uniform exploration in a finite subset of the $N$-body phase space composed on all those microscopic configurations where particles are trapped by the system gravitational field.

The previous hypothesis relies on dynamical foundations. Besides collisions, there is another general relaxation mechanism that explains a strong chaoticity of $N$-body nonlinear Hamiltonian dynamics, and hence, the possible justification of its ergodicity and mixing \cite{GallavottiCohen}. As already shown by Pettini and co-workers in the framework of geometrical description of Hamiltonian chaos \cite{Pettini,Pettini2}, systems with bounded motions in the configurational space can undergo the mechanism of \emph{parametric resonance}. Numerical studies in the framework of astrophysical systems have revealed that these systems exhibit strong chaotic properties even when collision events are practically nonexistent \cite{Pettini3}, which supports application of conventional statistical arguments even for collisionless stellar systems such as elliptical galaxies \cite{Cipriani}. An isolated astrophysical system is not able to satisfy ergodicity because of the accessible $N$-body phase space is unbound. Ergodicity is restored if the system is somehow forced to remain trapped inside a finite region of the $N$-body phase space. This requirement is achieved when the system is enclosed by a container with impenetrable walls, which provides here a \emph{spacial regularization} for the statistical description (thus avoiding the known long-range divergence of gravitation). Licitness of Antonov isothermal model relies on this argument. Alternatively, an effective confinement is also achieved by the incidence evaporation in presence of an external gravitational field, which introduces the restriction:
\begin{equation}\label{energ.rest}
\varepsilon(\mathbf{r},\mathbf{p})=\mathbf{p}^{2}/2m+m\varphi(\mathbf{r})<\varepsilon_{c}<0.
\end{equation}
This constraint introduces as a sort of \emph{energetic regularization} for the statistical description, where ergodicity is partially restored in a form of quasi-ergodicity. This means that the system will evolve along a certain quasi-stationary regime under these conditions, where its total energy and number of particles behave as quasi-conserved quantities.

\begin{figure}[tbp]
\begin{flushright}
\includegraphics[width=4.0in]{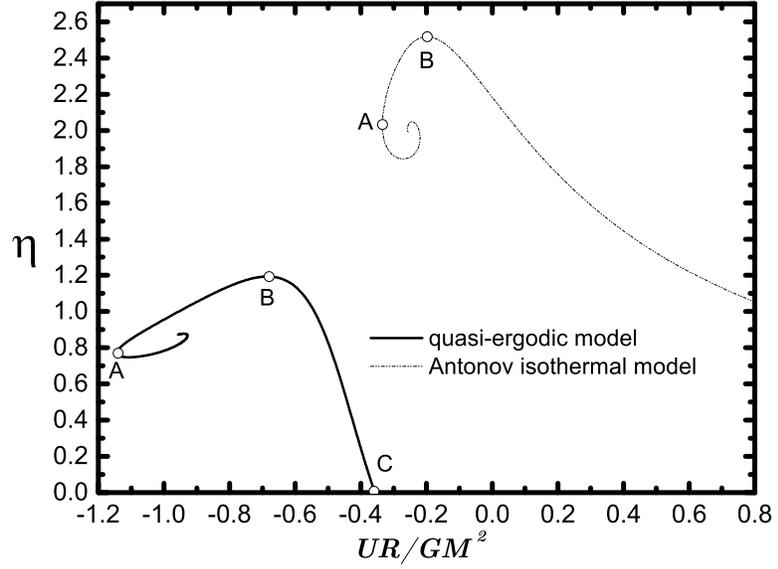}
\end{flushright}
\caption{Comparison between the caloric curves of the quasi-ergodic model and the Antonov isothermal model for dimensionless variables $u=UR/GM^{2}$ and $\eta=GMm/RkT$. Here, it was employed the same value for the radius of spherical container and the tidal radius. After \cite{Vel.QEM}.}\label{QEM.caloric.eps}
\end{figure}

Statistical description starts from the calculation of the number of microscopic configurations:
\begin{equation}\label{QEM.Integral}
\Omega\left[U,N|\varepsilon_{c}\right]=\frac{1}{N!}\int_{\mathcal{P}_{R}}\delta\left[U-H_{N}(\mathbf{r},\mathbf{p})\right]\prod^{N}_{i=1} \frac{d^{3}\mathbf{r}_{i}d^{3}\mathbf{p}_{i}}{\left(2\pi\hbar\right)^{3}},
\end{equation}
which is employed to obtain the microcanonical entropy of this system $S=k\ln W$. Here, $\mathcal{P}_{R}$ represents the subset of $N$-body phase space that fulfils the regularization constraint (\ref{energ.rest}). Hamiltonian $H_{N}(\mathbf{r},\mathbf{p})$ of the astrophysical system obeys the mathematical form (\ref{sgg}). The potential field $w_{c}(\mathbf{r})$ of the external forces now extracts all those particles whose individual mechanical energy $\varepsilon(\mathbf{r},\mathbf{p})$ overcomes the threshold energy $\varepsilon_{c}$. For the sake of simplicity, the mathematical form of this external field is idealized as follows:
\begin{equation}
w_{c}(\mathbf{r})=\left\{
\begin{array}{cc}
0 & \mbox{if } \left|\mathbf{r}\right|<R_{t}\\
-\infty & \mbox{otherwise,} \\
\end{array}
\right.
\end{equation}
where $R_{t}=-GMm/\varepsilon_{c}$ is the tidal radius. Formally, this external field mimics a container with a \emph{perfect absorbing walls}: every particle that reaches the boundary of this region is immediately extracted out the system. Integral (\ref{QEM.Integral}) can be estimated considering the continuum approximation and the steepest descend method. In general, the quasi-ergodic model is able to access the same qualitative features derived from truncated $\gamma$-exponential models, such as distributions profiles with isothermal cores and polytropic haloes; states with negative heat capacities, and the existence of collective behaviors such as gravitational collapse and evaporation disruption. A comparison between caloric curves of Antonov isothermal model and quasi-ergodic model is shown in figure \ref{QEM.caloric.eps}. A distinguishing feature of this model is the presence of driving function $C(\mathbf{r})$ in the distribution profile of this model:
\begin{equation}\label{qem.pdf}
f_{qs}(\mathbf{r},\mathbf{p})=Ae^{C(\mathbf{r})}e^{\beta\left[\varepsilon_{c}-\varepsilon(\mathbf{r},\mathbf{p})\right]}
\Theta\left[\varepsilon_{c}-\varepsilon(\mathbf{r},\mathbf{p})\right],
\end{equation}
which is obtained by self-consistence:
\begin{equation}
C(\mathbf{r})=-\int \beta\frac{G m^{2}}{\left|\mathbf{r}-\mathbf{r}'\right|}A e^{C(\mathbf{r}')}2\sqrt{\frac{\Phi(\mathbf{r}')}{\pi}}d^{3}\mathbf{r}',
\end{equation}
with $\Phi(\mathbf{r})=\beta \left[\varepsilon_{c}-m\varphi (\mathbf{r})\right]$ being the dimensionless potential. All previous models include the incidence of evaporation as a power-law truncation of Boltzmann-Gibbs distribution at the threshold energy. However, the present thermo-statistical arguments describe an additional and nontrivial consequence of evaporation: a nonuniform redistribution of particles that arises as a long-range effect of gravitation.

\section{Astrophysical systems in thermodynamic interaction}

\subsection{Non-extensive character and long-range correlations}\label{non-extensive}

The incidence of \emph{surface effects} in the framework of extensive systems can be neglected in comparison with bulk effects whenever the system size (e.g.: the number of particles $N$ or the volume $V$) is very large. For example, the boiling temperature of water can depend on the environmental pressure $p$, but it does not depend on the volume of container or its shape whenever one considers the thermodynamic limit:
\begin{equation}\label{ext.TL}
N\rightarrow\infty:\mbox{ where }U/N\mbox{ and }V/N\mbox{ are kept constant}.
\end{equation}
In contrast, the microcanonical caloric curves shown in figures \ref{Antonov.eps} and \ref{King.eps} actually depend on the radius $R$ of container or the tidal radius $R_{t}$. It is possible to verify that thermodynamical behavior of Antonov isothermal model depends both on the container shape and its linear dimensions \cite{Vega1}. Moreover, King model and its generalization, the truncated $\gamma$-exponential model, illustrate that the incidence of evaporation produces a considerable affectation in thermodynamics of astrophysical systems, and in extreme cases, it is able to suppress the occurrence of gravitational collapse as the situation described by Plummer model.

This remarkable behavior is associated with the \emph{long-range character} of gravitational interaction. Any external influence, even acting on the boundary or a small portion of the system, will affect the system as a whole. In the framework of extensive systems, such a \emph{sensibility} to the external influence is observed during the occurrence of a critical phenomenon because of the divergence of correlation length $\xi$ \cite{Reichl}. Roughly speaking, systems with long-range interactions, as the astrophysical ones, behave as \emph{they permanently were in a critical state}, although with a less dramatic character. For example, the statistical correlations among particles in conventional extensive systems follow an exponential behavior as $C(\mathbf{r}_{i},\mathbf{r}_{j})\propto\exp\left[-\left|\mathbf{r}_{i}-\mathbf{r}_{j}\right|/\xi\right]$, where the correlation length $\xi$ is comparable to the system linear size near critical point \cite{Goldenfeld}. For the case of astrophysical systems, the two-point correlations are replaced by a power-law dependence as $C(\mathbf{r}_{i},\mathbf{r}_{j})\propto1/\left|\mathbf{r}_{i}-\mathbf{r}_{j}\right|^{\alpha}$, although one should observe a stronger dependence during the occurrence of gravitational collapse. Some studies in the past revealed that the large-scale distribution of galaxies in the universe seems to follow this type of power-law with exponent $\alpha\simeq 1$.$81$ \cite{Luo}, and presumably, other more complex behaviors are relevant in different length scales \cite{Labini}.

The ubiquitous existence of long-range correlations in astrophysical situations is a direct consequence of the long-range character of gravitation. Of course, some macroscopic behaviors can be related to the fact that Newtonian gravitation $w(\mathbf{r},\mathbf{r}')=-Gmm'/\left|\mathbf{r}-\mathbf{r}'\right|$ is intrinsically \emph{scale-free interaction}. A purely classical self-gravitating gas of non-relativistic point particles interacting via Newtonian gravitation, as the one described by Hamiltonian (\ref{sgg}) with external potential $w_{c}(\mathbf{r})=0$, has no characteristic energy or linear dimension. In fact, the characteristic lengths and energies for most of astrophysical models reviewed in this work are determined by the external conditions. A remarkable exception here is the Plummer model, where the characteristic linear dimension of particles distribution (\ref{Plummer.length}) is determined from its own total energy $U$ and mass $M$. This astrophysical model exhibits what it could be regarded as a \emph{scale-free thermodynamic behavior}. The absence of characteristic lengths is the reason why this model cannot describe collective behaviors like gravitational collapse or evaporation disruption.

\subsection{Objections against Lynden-Bell and Wood explanation of gravothermal catastrophe: non-applicability of conventional notions of separability and additivity in astrophysics}

Lynden-Bell and Wood \cite{Lynden-Bell1} employed the notion of negative heat capacity to explain gravothermal collapse of Antonov isothermal model \cite{Antonov}. Their arguments were based on the application of stability condition (\ref{Thirring}) considering the positive heat capacity of \emph{outer regions} of the system and the negative heat capacity of its \emph{inner ones}\footnote{Although Thirring derived the stability criterium (\ref{Thirring}) two years after publication Lynden-Bell and Wood, these last authors were aware about its existence. According to their own words: \emph{... if the outer parts have a specific heat smaller than the magnitude of the negative specific heat of the inner parts, then the outer temperature will increase faster than the inner one. It will therefore catch up and a possible equilibrium state will be achieved. If however the outer parts are too extensive their specific heat will be large. Conductive transfer of heat from the central parts will then raise the high central temperature faster than it raises the lower temperature of the outer parts. This is what we call the \textbf{gravothermal catastrophe} ...}}. Although their reasonings were quite convincing, a re-examination of this situation clarifies that this stability criterion cannot be naively applied here. The derivation of Thirring inequality (\ref{Thirring}) involves two strong conditions: (i) the \emph{separability} of a closed system into two independent subsystems, and (ii) the \emph{additivity} of the total energy due to the existence of a thermal contact. None of these conditions is met for the astrophysical situation considered by these authors.

A single astrophysical structure (e.g.: a star or a stars cluster) cannot trivially divided into subsystems in a way that its total entropy could be expressed as sum of respective entropies of these subsystems. Gravitation is a long-range force, and hence, the constituents a given astrophysical structure are strongly correlated among them. In fact, it is \emph{impossible to modify a part of the same one without affecting the whole astrophysical structure}. This very fact is evidenced in each astrophysical model previously considered, including the proper Antonov isothermal model. However, the additivity of the total entropy arises as a good approximation when a closed system is already divided into a very small number of astrophysical structures., e.g.: a binary star system, a star cluster and its host galaxy, etc. Such an effective separability is implemented in terms of \emph{internal} and \emph{collective degrees of freedom}.

The separability of a given closed system into subsystems is a pre-requisite for the existence of \emph{additivity} for a given physical quantity. However, the separability itself does not guarantee the additivity. The additivity of the energy under the \emph{thermal contact} represents a good approximation in the framework of large extensive systems of everyday applications since they are driven by \emph{short-range interactions}. The constituents of each system that are affected by the presence of the other one are mostly located near their surfaces, so that, the total interaction energy associated to thermal contact can be regarded as a \emph{surface effect}, which is negligible in comparison to the bulk internal energies of each system. In contrast, the interaction among astrophysical systems involve all their constituting particles because of gravitation is a long-range interaction. The associated interaction energy among astrophysical systems represent \emph{bulk effects}, so that, the additivity of the total energy does not hold in general\footnote{The additivity of energy is trivially restored when these astrophysical structures are displaced far away from each other. Since their mutual interactions become negligible, each system is put into energetic isolation.}. A better understanding about how separability and additivity are extended to astrophysical situations will be analyzed in the subsection \ref{gen.stat.cond}.

To avoid any misunderstanding of potential readers, it is important to remark that gravitational collapse is a type of thermodynamic instability that is present for most of astrophysical systems in any scale. What is wrong here is the explanation of this phenomenon in terms of separability of a given astrophysical system into independent subsystems. Gravitational collapse is related to presence of negative heat capacities since both behaviors arise as direct consequences of radial dependence $-1/r$ of gravitational interaction (the called short-range divergence). However, one can advertise an additional relationship analyzing the energy dependence of specific heat $c=C/Nk$ shown in inset panel of figure \ref{Antonov.eps}. The critical energy of gravitational collapse $u_{A}$ also coincides with the \emph{eventual vanishing of the heat capacity}. Once again, this mathematical behavior may suggests a misleading impression that heat capacity becomes zero as if the system would be decomposed of separable outer and inner parts with the appropriate heat capacity. However, this behavior is more analogue to the vanishing of heat capacity of quantum gases in the zeroth temperature limit, where their internal degrees of freedom (e.g.: translational, vibrational and rotational) are \emph{effectively frozen}. The astrophysical case is different since the temperature grows when the system approaches the point of gravitational collapse. The system as a whole undergoes at this critical point a collective fall under its own gravitational field, which leads to a sudden formation of a \emph{core-halo structure} and the transformation of potential energy into kinetic. This \emph{collective process of self-organization} is necessarily accompanied with a growth of the statistical correlations among the particles. The number degrees of freedom is effectively reduced here because of they lose their statistical independence, which would explain the sudden reduction of heat capacity per particles.

From a thermo-statistical viewpoint, the gravitational collapse can be related to a new type of collective phenomena: the called \emph{discontinuous microcanonical phase transitions} \cite{Antoni}. This phase transition involves an exotic type of \emph{ergodicity breaking} \cite{Reichl}, where the system is dynamically trapped in one among two or more different stable configurations that exhibit the same value of energy (stable configurations with different internal conditions for temperature, central density, pressure, etc.). Antonov isothermal model does not show this type of phase transitions since it disregards the \emph{underlying physics} in the post-collapse phase. As already discussed in the literature \cite{CasettiMK,Chavanis2002,Chavanis2004}, the system can find a new branch of stable configurations after gravothermal collapse if the \emph{quantum behavior} of particles is taken into account, or even, when the particles exhibit an internal structure like a small \emph{hard-core} or \emph{soft-core} radius.

\subsection{Open astrophysical systems}

Thermo-statistics of astrophysical systems is hallmarked by the occurrence of gravitational collapse at low energies and the existence of states with negative heat capacities. Among conventional ensembles of equilibrium statistical mechanics, the microcanonical ensemble is the only able to describe these macroscopic behaviors. This statistical ensemble, however, presupposes that the system of interest is put into energetic isolation, which is a situation that does not correspond to actual conditions of astrophysical systems in Nature. Most of real astrophysical systems belong to the class of \emph{open systems}, that is, systems that are able to interchange energy and matter with other astrophysical systems (e.g.: stars are part of multiple stars systems; stars clusters are under gravitational influence of their host galaxy; galaxies themselves undergo the gravitational influence of other galaxies in a galactic cluster, etc.).

Usually, open systems in equilibrium are described within canonical or gran canonical ensembles. Although these statistical ensembles are employed in the literature to analyze astrophysical problems, their consideration is merely a theoretical exercise without practical support. How does one justify the existence of a \emph{thermal contact} between an astrophysical system and a Gibbs thermostat? Besides, what physical system meets the necessary requirements to be considered as a good approximation of a Gibbs thermostat in the astrophysical scenario? From a purely theoretical perspective, one can enclose the system inside a container whose walls are kept at constant temperature \cite{Antonov}, or immerse the astrophysical system into an infinite medium with constant temperature \cite{Chavanis.etal}, among others. However, all these arguments are based on our practical experience with short-range extensive systems, which is irrelevant to deal with astrophysical problems. Astrophysical systems in Nature interact with other astrophysical systems only, which are necessarily coupled among them throughout gravitation. Clearly, these situations do not correspond to the ones considered by conventional statistical ensembles. The challenge here is the derivation of more suitable statistical ensembles, as example, astrophysical counterparts of canonical and gran canonical ensembles.

Why is this subject so important? As already emphasized in section \ref{NHC}, the thermodynamic behavior of a given macroscopic system actually depends on the external conditions that were impose to. This situation is known in the literature as \emph{inequivalence of ensembles} \cite{Bouchet,Barre}. However, the importance of this phenomenon in astrophysical problems is more significant. The long-range character of gravitation enables that any change in the external conditions will modify the \emph{system bulk properties}, such as the equations of state, the response functions, and even, the occurrence of phase transitions. Accordingly, a realistic description of each open astrophysical system requires a full specification of its associated external conditions.

\subsection{Other paradox concerning to negative heat capacities}

The possible generalization of conventional statistical ensembles for astrophysical scenario requires to face the following question: Can two or more astrophysical systems in mutual interaction, each one with a negative heat capacity, remain in equilibrium (or at least, in a quasi-stationary situation)? According to Thirring stability conditions (see in subsection \ref{Thirring}), two systems with negative heat capacities cannot remain in equilibrium. However, a negative answer to the previous question could be in contradiction with astrophysical observations. Most of stars in Universe are part of multiple star systems, binary stars as example \cite{Heintz}. Strictly speaking, stars are thermodynamic systems with negative heat capacities that are fully able to interchange energy by mean of their gravitational interaction, as example, they undergo heating by tidal interactions \cite{Zahn}. If one star gains energy from its companion in a binary system, its temperature decreases while the companion temperature increases, a situation that in some way should speed-up the rate of energy transference. Such a process will unavoidably lead to the gravitational collapse of the star that looses energy and the evaporative disruption of its companion. In principle, this picture could be a regarded as a licit description of final stage of evolution of these systems, but binary stars should remain stable the major part of their life. Notice that the stability analysis of these systems concerns both the dynamic stability and the processes of mass transference \cite{Paczynski,Kopal}.

Typically, it is regarded that these stellar systems evolve as \emph{independent thermodynamic systems}, which is obviously a simplifying consideration. This situation could be explained by two reasons: (i) the rate of energy transference between the stars in a binary system is extremely low, or (ii) there is something \emph{wrong} with application of Thirring stability conditions in astrophysical situations. This paradox situation is partially solved from the fact that the starting considerations of Thirring stability criteria (\ref{Thirring}) do not apply for the case of astrophysical systems. Two separable astrophysical systems in mutual interaction could support an additive total entropy, $S_{T}\simeq S_{A}+S_{B}$. However, their total energy $U_{T}$ cannot be expressed as the sum of their respective internal energies as $U_{T}\neq U_{A}+U_{B}$ because of the long-range character of gravitation. I think that thermodynamic stability could be possible under certain conditions, but it is necessary to perform a re-examination of Thirring stability analysis for astrophysical situations.

\subsection{Thermodynamic equilibrium or quasi-equilibrium?}

Before to enter to analyze the conditions of thermodynamic stability, let us discuss about the relevance of \emph{thermodynamic equilibrium} for real astrophysical systems. As discussed elsewhere, this is a suitable idealization that is assumed to face the study of conventional systems of everyday applications of thermodynamics. Its existence is postulated in order to introduce the temperature notion, which is keystone that supports other notions, methods and laws of thermodynamics. Conceptually, the existence of thermodynamic equilibrium plays a similar role of the \emph{law of inertia} in classical mechanics. The later one is employed as a pre-requisite to introduce the notion of \emph{force} as a quantitative measure of interactions, while temperature differences (or gradients) behave as \emph{generalized thermodynamic forces} that drive the fluxes of energy and heat among the macroscopic systems. In an analogous way that a real system cannot fully be isolated from the incidence of external forces, the systems in Nature cannot be found in a rigorous thermodynamic equilibrium. At most, they exhibit a \emph{quasi-stationary evolution} under certain background conditions, where the rate of change of their macroscopic observables is sufficiently slow during the lapse time of interest. For practical purposes, such a very slow evolution of macroscopic behavior is sufficient to speak about a sort of \emph{quasi-equilibrium}. Our practical experience evidences that methods of thermodynamics and statistical mechanics can be applied to predict the macroscopic properties of systems in quasi-stationary conditions.

At first glance, the study of real astrophysical systems present serious difficulties. Not only their thermodynamic properties are different, but the external background conditions that operate in a given astrophysical situation are \emph{non-controllable}. This very fact contrasts with the case of conventional extensive systems that are studied in our laboratories. The usage of idealizations such as containers with impenetrable walls or non-conducting boundaries is highly nonphysical in the astrophysical scenario. Consequently, the incidence of evaporation and the energy dissipation will always play a significant role here. Does this mean that we should renounce the use of all these thermodynamic idealizations in the case of astrophysical systems? On the contrary, concepts and methods of thermodynamics should be adapted to this physical scenario. In particular, the notion of thermodynamic equilibrium should be replaced by a quasi-equilibrium, which is also relevant to study traditional systems of everyday practice.

\begin{figure}[tbp]
\begin{flushright}
\includegraphics[width=7.0in]{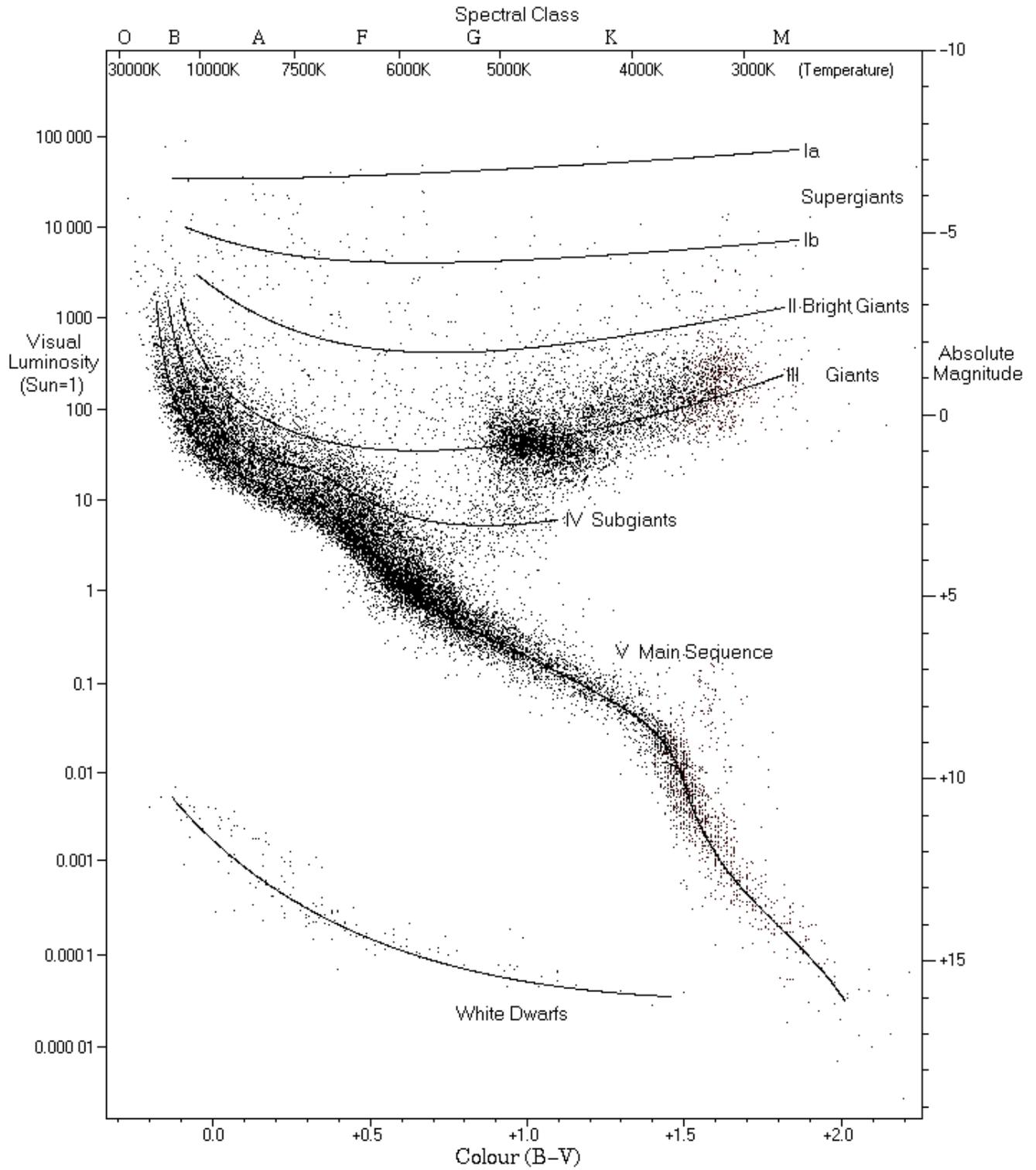}
\end{flushright}
\caption{Hertzsprung-Russell diagram with 22000 stars plotted from the Hipparcos Catalogue and 1000 from the Gliese Catalogue of nearby stars. After \cite{Richard}.}\label{hr.eps}
\end{figure}

Is the notion of quasi-equilibrium relevant for real astrophysical systems? Judge yourself. Stars are open astrophysical systems that continuously release energy and matter to the space. However, they reach a quasi-stationary evolution with predictable regularities. The best example of these regularities is the temperature-luminosity (or color-magnitude) relation represented by the known Hertzsprung-Russell diagrams \cite{Rosenberg}, as the one shown figure \ref{hr.eps}. Moreover, the effective surface temperature of stars and the spectrum of solar radiation can also described with a good accuracy by the laws of black-body thermal radiation. Other example is the case of globular clusters, whose distribution profiles and evolution can be explained with the help of simple statistical models, like King models reviewed in this work \cite{King}. Galaxies them-selves suffer from gravitational interaction of other galaxies, and they presupposed evolve without the incidence of collisions. Surprisingly, the brightness profiles of galaxies exhibit simple regularities such as Vaucouleurs and Sersic profiles \cite{Vaucouleurs,Sersic}, while their associated dark matter haloes also follow other predictable regularities \cite{Kravtsov}. Finally, the cosmic microwave background radiation also follows the law of blackbody thermal radiation regardless the Universe itself is not in thermodynamic equilibrium \cite{PenziasWilson,Fixsen}, but it undergoes an accelerate expansion.

The existence of the above regularities is explained by the incidence of \emph{relaxation mechanisms} that lead all these astrophysical situations to quasi-stationary conditions. Of course, a Maxwellian velocity distribution does not apply for a real astrophysical system. However, its applicability also fails when quantum properties of matter turn relevant. In principle, the incidence of evaporation should be seem as another \emph{modifying factor} for this distribution. Evaporation is just a relaxation mechanism in this context. For the case of stars clusters, those stars with sufficient energy to escape will be removed by the incidence of external tidal forces of a nearby galaxy, so that, the remaining stars will be more bound among them, which restrains further escapes. Inverse temperature parameter $\beta=1/kT$ of Maxwell distribution represents the scaling parameter of velocities, which is the same role of parameter $\beta$ in lower-isothermal Michie-King distribution (\ref{MK.pdf}) or distribution profile (\ref{qem.pdf}) of quasi-ergodic model. The notion of temperature is not associated with a distribution with concrete mathematical form. It is the quantity that characterizes the energy interchange among macroscopic systems in thermodynamic equilibrium. In principle, an effective temperature could be introduced to characterize quasi-stationary situations. Although quasi-stationary profile (\ref{qem.pdf}) of quasi-ergodic model is non-Maxwellian, its associated inverse temperature parameter $\beta=1/kT$ is obtained from the microcanonical relation (\ref{mic.def}), which also enters in a sort of \emph{Legrendre transformation} between thermodynamic potentials (see in Ref. \cite{Vel.QEM} for further details).

Anyway, the realistic possibility that an astrophysical situation is found in a quasi-stationary regime (or quasi-equilibrium) requires the reconsideration of some definitions and terms. Let us hereinafter refer by \emph{thermodynamic stability in conditions of quasi-equilibrium} to the persistence or the tendency of a system to remain its evolution along this quasi-stationary regime. On the contrary case, one can speak about thermodynamic instability. It is expected that a system in conditions of thermodynamic instability experiences a much faster evolution, if possible, towards a new regime of quasi-stationarity. Everywhere in this work, when one refers to \emph{thermodynamic equilibrium} or \emph{stability} for astrophysical situations, it should be implicitly understood that they refer to \emph{systems in conditions of quasi-stationarity}.

A particular illustration about the occurrence of this type of processes in Nature is the proper Hertzsprung-Russell diagram shown in figure \ref{hr.eps}, which reveals the different stages of stellar evolution \cite{Kaufmann}. Most of stars, including our Sun, are evolving along the called main-sequence the major part of their life, where the energy is supplied by the fusion of hydrogen atoms at the star core. With the eventual preponderance of helium atoms in that region, stars like the Sun begin to fuse hydrogen along a layer surrounding the core. This process causes a gradual growth of the star size, which abandons thus the main-sequence and passing through the subgiant stage until it reaches the red giant phase. Once a star like the Sun has exhausted its nuclear fuel, the same one will experience a fast gravitational collapse. Along the occurrence of this process, the star core becomes a dense white dwarf and the outer regions are expelled as a planetary nebula. Stars with larger masses can explode in a supernova, while their cores collapse to form an extremely dense neutron star, or even, a black hole. The total mass is the most important quantity in the stellar stability and evolution. However, this process also depends on other factors such as the chemical composition, as well as whether or not they are under the external influence of neighboring systems. Most of stars in Universe are part of multiple star systems. The evolution of a star in a binary system, as example, could be affected by the separation distance, the total mass and the stage of evolution of its companion. Even, the occurrence of close encounters among stars could play an important role in certain situations \cite{Leonard}.

\subsection{Thermal equilibrium conditions revisited}\label{gen.stat.cond}

\begin{figure}[tbp]
\centering
\includegraphics[width=2.5in]{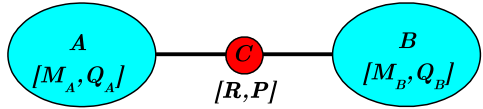}
\caption{Two separable astrophysical systems A and B are coupled throughout its collective degrees of freedom $[\mathbf{R},\mathbf{P}]$, which behaves here as a third system C (one-body system). The quadrupole-orbit coupling in (\ref{quadrupole}) is responsible of a \emph{tidal deformation} of these systems, as well as the existence of finite energy threshold $\varepsilon_{c}$ for the escape of particles of each system from the influence of its own gravitational field and be trapped by the one of the other system. In principle, the heat transference between these systems could be addressed considering a sort of \emph{Langevin equation} that describes the dynamics of low dimensional system C under the influence of two \emph{heat baths} with finite negative heat capacities. It is noteworthy that this situation is the basis for the plausible introduction of \emph{astrophysical counterparts} of the known canonical and gran canonical statistical ensembles considering the limit $M_{A}\ll M_{B}$. Such a picture could be relevant for describing the thermodynamic interaction among a gralaxy and its stars clusters in astrophysics.}\label{CoupledBaths.eps}
\end{figure}

Let us consider two astrophysical systems A and B in mutual gravitational interaction. This situation can be described in terms of their respective internal and collective degrees of freedom. For each system, the internal degrees of freedom are the positions and their conjugated momenta $\left\{\mathbf{r}_{i},\mathbf{p}_{i}\right\}$ relative to the center of mass, while the collective degrees of freedom $(\mathbf{R},\mathbf{P})$ refer to the relative separation vector of the centers of mass and its conjugated momentum, respectively. Denoting by $H_{A}$ and $H_{B}$ the contributions of internal degrees of freedom, the Hamiltonian $H$ of this closed system can be expressed as follows:
\begin{equation}
H=H_{A}+H_{B}+W_{AB},
\end{equation}
where $H_{\alpha}$ with subindex $\alpha=(A,B)$ denotes the contribution of internal degrees of freedom of each system:
\begin{equation}
H_{\alpha}=\sum_{i_{\alpha}}\frac{1}{2m_{i_{\alpha}}}\mathbf{p}^{2}_{i_{\alpha}}
-\sum_{i_{\alpha}<j_{\alpha}}\frac{Gm_{i_{\alpha}}m_{j_{\alpha}}}{\left|\mathbf{r}_{i_{\alpha}}-\mathbf{r}_{j_{\alpha}}\right|},
\end{equation}
while the term $W_{AB}$:
\begin{equation}
W_{AB}=\frac{1}{2\mu_{AB}}\mathbf{P}^{2}-\sum_{i_{A}i_{B}}\frac{Gm_{i_{B}}m_{i_{A}}}{\left|\mathbf{R}+\mathbf{r}_{i_{A}}-\mathbf{r}_{i_{B}}\right|}
\end{equation}
is the interaction energy that involves the contribution of collective motions and its couplings with particle positions $\left\{\mathbf{r}_{i_{A}}\right\}$ and $\left\{\mathbf{r}_{i_{B}}\right\}$ of each system. As usual, the kinetic energy of collective motions is characterized by the reduced mass $\mu_{AB}=M_{A}M_{B}/\left(M_{A}+M_{B}\right)$, which is expressed in terms of the total masses $M_{A}$ and $M_{B}$ of each system. The internal degrees of freedom must satisfy the additional constraints for the center of mass:
\begin{equation}
\mathbf{P}_{\alpha}=\sum_{i_{\alpha}}\mathbf{p}_{i_{\alpha}}=0\mbox{ and }\mathbf{R}_{\alpha}=\frac{1}{M_{\alpha}}\sum_{i_{\alpha}}m_{i_{\alpha}}\mathbf{r}_{i_{\alpha}}=0.
\end{equation}
A suitable expression for $W_{AB}$ that describes the lowest coupling among collective and internal degrees of freedom is the \emph{quadrupole approximation}:
\begin{equation}\label{quadrupole}
W_{AB}\simeq\frac{1}{2\mu_{AB}}\mathbf{P}^{2}-\frac{GM_{A}M_{B}}{\left|\mathbf{R}\right|}
-\frac{G}{2\left|\mathbf{R}\right|^{3}}\left[M_{A}Q_{B}+M_{B}Q_{A}\right].
\end{equation}
One has adopted here the notations $Q\equiv Q^{\alpha\beta}n_{\alpha}n_{\beta}$, with $n_{\alpha}$ being the Cartesian components of unitary vector $\boldsymbol{n}=\mathbf{R}/\left| \mathbf{R}\right|$. Finally, $Q^{\alpha\beta}$ is the quadrupolar tensor of each system:
\begin{equation}
Q^{\alpha\beta}=\sum_{i}m_{i}\left[3r_{i}^{\alpha}r_{i}^{\beta}-\delta^{\alpha\beta}\mathbf{r}^{2}_{i}\right].
\end{equation}
Second terms in (\ref{quadrupole}) represent the called \emph{quadrupole-orbit coupling}, which is schematically represented in figure \ref{CoupledBaths.eps}. Collective motions behave as \emph{a third astrophysical system C} with low dimensionality (one-body system). Expression (\ref{quadrupole}) evidences that the interaction energy $W_{AB}$ is a \emph{bulk effect}, which is determined by quantities that characterize the astrophysical systems as a whole: their respective total masses $\left[M_{A},M_{B}\right]$, quadrupole tensors $\left[Q_{A},Q_{B}\right]$ and collective degrees of freedom $(\mathbf{R},\mathbf{P})$. Accordingly, the energy interchange among internal degrees of freedom of each system is driven by the collective motions, which is a thermodynamic mechanism fully different of those ones that operate in extensive systems.

Since the characteristic timescale of collective motions $\tau_{C}$ significantly greater than the characteristic timescales of internal degrees of freedom, $\tau_{C}\gg (\tau_{A},\tau_{B})$, one can consider a sort of \emph{adiabatic approximation}\footnote{Adiabatic approximation enters in quantum-mechanics description of atoms and molecules, where it is known as \emph{Born-Oppenheimer approximation}. Essentially, it assumes that the motions of atomic nuclei and electrons can be separated, which is expressed by a factorization of wavefunction of a molecule into its electronic and nuclear components $\Psi_{molecular}\simeq \Psi_{nuclear}\Psi_{electronic}$. This approximation is justified because of electrons evolve much faster than the nuclei due to the significant difference in their masses. For the astrophysical case, the adiabatic approximation is justified by the small values of the ratios $\tau_{A,B}/\tau_{C}\simeq\left[3(M_{A}+M_{B})/R^{3}\rho_{A,B}\right]^{1/2}\ll 1$, where $\rho_{A,B}$ is a measure of central mass density of each system.}. Specifically, one can assume that the collective motions are practically \emph{frozen} when one is interested on the study of internal degrees of freedom, while the study the collective motions can be performed by restricting to the \emph{average dynamical influence} of internal degrees of freedom. In principle, this approximation could be employed to analyze aspects concerning to equilibrium (or quasi-equilibrium), such as stability conditions and the possible generalized statistical ensembles, as well as non-equilibrium features like the \emph{heat transport}. For example, one could consider the dynamical evolution of collective degrees of freedom as a low dimensional system that is coupled with two heat baths (the internal degrees of freedom of each system). This is a typical arrangement employed in non-equilibrium thermodynamics, except for the fact that \emph{these heat baths are actually astrophysical systems with negative heat capacities}.

According to expression (\ref{quadrupole}), the internal and collective degrees of freedom become almost independent when the separation $\left |\mathbf{R}\right|$ is large enough. The entropy $S_{T}$ of the closed system can be expressed as follows:
\begin{equation}
S_{T}=S_{A}+S_{B}+S_{C}+S_{ABC},
\end{equation}
where $(S_{A},S_{B},S_{C})$ are the contributions of internal and collective degrees of freedom, while $S_{ABC}<0$ describes a reduction due to their existing correlations associated with the presence of quadrupole terms in (\ref{quadrupole}). Since the number of internal degrees of freedom of each system are considerably much larger than the one of collective motions, it is natural to assume the application of the following inequalities\footnote{The entropy of any physical system is highly dependent on the number of particles $N$, while the internal energy $U$ appears in a logarithmic dependence. This fact can be verified in entropy estimation (\ref{mic.entropy}).}:
\begin{equation}
\left|S_{ABC}\right|<\left|S_{C}\right|\ll\min\left[\left|S_{A}\right|,\left|S_{B}\right|\right]
\end{equation}
whenever the \emph{separability} of astrophysical systems A and B is a licit assumption. Accordingly, the contributions $S_{C}$ and $S_{ABC}$ can be neglected with a great accuracy, and the total entropy $S_{T}$ of the closed system obeys additivity:
\begin{equation}
S_{T}\simeq S_{A}+S_{B}.
\end{equation}
Of course, this separability fails in the limit $\left |\mathbf{R}\right|\rightarrow 0$, where it takes places the merge of these astrophysical systems.

The separability of these astrophysical system can guarantee the additivity of the total entropy $S_{T}$, but additivity of the total energy $U_{T}$ is not longer a licit approximation. In the framework of adiabatic approximation, the statistical average of the interaction energy $W_{AB}$ can be regarded as a function on the internal energies $\left(U_{A},U_{B}\right)$ of each system, where the collective degrees of freedom $(\mathbf{R},\mathbf{P})$ enter as external control parameters. The simplest situation is that the collective motions nearly evolve along a circular orbit with the mean separation distance $D$. Considering the ansatz:
\begin{equation}\label{ansatz}
\left\langle W_{AB}\right\rangle=W\left(U_{A},U_{B}|D\right),
\end{equation}
the additivity of the total energy $U_{T}=U_{A}+U_{B}$ is now replaced by the following constraint:
\begin{equation}\label{energ.constraint}
U_{T}=U_{A}+U_{B}+W\left(U_{A},U_{B}|D\right).
\end{equation}
This expression describes a \emph{non-linear energy interchange} between the astrophysical systems, where the amount of energy $\Delta U_{A}$ transferred by the system A differs from the amount of energy $-\Delta U_{B}$ absorbed by the system B, and vice versa. A fraction of this energy is employed to change the state of collective motions, and hence, \emph{the incidence of this effect modifies the conventional thermal equilibrium conditions}.

Combining the differential expression:
\begin{equation}
dU_{T}=\left[1+\left(\frac{\partial W}{\partial U_{A}}\right)\right]dU_{A}+\left[1+\left(\frac{\partial W}{\partial U_{B}}\right)\right]dU_{B}=0
\end{equation}
and the total derivative of the entropy $S_{T}$ at its local maximum, one obtains:
\begin{equation}\label{astro.EqCond}
\frac{\partial S_{T}}{\partial U_{A}}=0\Rightarrow \phi_{A}\frac{\partial S_{A}}{\partial U_{A}}=\phi_{B}\frac{\partial S_{B}}{\partial U_{B}}\left(\equiv \eta\right),
\end{equation}
where the quantities $\phi_{\alpha}$ are defined as:
\begin{equation}\label{factors.coll}
\phi_{\alpha}=\left[1+\left(\frac{\partial W}{\partial U_{\alpha}}\right)\right]^{-1}.
\end{equation}
Formally, equation (\ref{astro.EqCond}) is a generalization of thermal equilibrium condition (\ref{stat.eq}). Since $\beta_{\alpha}=\partial S_{\alpha}/\partial U_{\alpha}$ is the \emph{microcanonical inverse temperature} of each system, the quantity $\eta_{\alpha}=\phi_{\alpha} \beta_{\alpha}$ is a sort of \emph{effective inverse temperature} that accounts for the influence of collective motions. According to the present analysis, the ordinary equalization of microcanonical inverse temperatures does not hold in astrophysical situations. Nevertheless, this property can be generalized in terms of their effective inverse temperatures, equation (\ref{astro.EqCond}). Considering an analogous procedure, one also obtains the stability condition:
\begin{equation}\label{ineq.Thirring.Gen}
\frac{\partial^{2}S_{T}}{\partial U^{2}_{A}}<0\Rightarrow \phi_{A}\left[\frac{C_{A}+C_{B}}{C_{A}C_{B}}-\frac{1}{\eta}\left(\frac{\partial\phi_{A}}{\partial U_{A}}+\frac{\partial\phi_{B}}{\partial U_{B}}\right)\right]>0,
\end{equation}
which generalizes inequality (\ref{stab}). Maximizing the total entropy $S_{T}$ considering the internal energy $U_{B}$ of the system B, the factor $\phi_{A}$ in (\ref{ineq.Thirring.Gen}) is replaced by $\phi_{B}$, which means that $\phi_{A}\phi_{B}>0$. According to this result, \emph{thermal equilibrium between two astrophysical systems crucially depends on the non-linear energetic interchange throughout the collective motions}. In particular, thermal equilibrium between systems with negative heat capacities could be possible whenever energetic interchange guarantees the inequality (\ref{ineq.Thirring.Gen}). Considering the limit $\phi_{\alpha}\rightarrow1$ when $D\rightarrow+\infty$, it is natural to expect that the positive character of these quantities for relevant astrophysical situations, $\phi_{\alpha}>0$. Apparently, thermal stability for systems with negative heat capacities could be possible if $\partial \phi_{\alpha}/\partial U_{\alpha}<0$. Unfortunately, one cannot go more further in this analysis without considering a concrete mathematical expression for the average interaction energy (\ref{ansatz}). Its form must be derived from a thermo-statistical model that includes the influence of quadrupole-orbit coupling (\ref{quadrupole}) on the thermodynamics of internal degrees of freedom.

\subsection{Roughly estimation of interaction energy $W_{AB}$}

Let us consider the co-moving reference frame that is located at the center of mass of one of the systems, as example, the system A. Let us also consider that the separation vector $\mathbf{R}$ between the centers of mass is fixed along $x$-axis of this co-moving reference frame. For the sake of simplicity, let us consider that the relative motion between the systems A and B is a circular orbit in the plane $x-y$, with radius $D$ and rotation frequency $\omega_{R}$:
\begin{equation}
\omega^{2}_{R}=\frac{G\left(M_{A}+M_{B}\right)}{D^{3}}.
\end{equation}
The system A is under the external influence of an \emph{effective tidal field} due to the quadrupole term in (\ref{quadrupole}) and the non-inertial centrifugal forces:
\begin{equation}\label{tidal.V}
W_{tidal}(\mathbf{r})=-\frac{1}{2}\omega^{2}_{R}\sum_{i}m_{i}\left[(1+2\xi_{A})x^{2}_{i}+(1-\xi_{A})y^{2}_{i}-\xi_{A}z^{2}_{i}\right],
\end{equation}
where $(x_{i},y_{i},z_{i})$ are the Cartesian components of particles positions $\mathbf{r}_{i}$, while $(\xi_{A},\xi_{B})$ are the auxiliary mass weights:
\begin{equation}
\xi_{A}=\frac{M_{B}}{M_{A}+M_{B}}\mbox{ and }\xi_{B}=\frac{M_{A}}{M_{A}+M_{B}}.
\end{equation}
The tidal potential (\ref{tidal.V}) should be combined with Hamiltonian contribution $H_{A}$ of internal degrees of freedom of the system A in order to develop a thermo-statistical model. Some features of this situation are described by King model \cite{King}, but this later one disregards the anisotropic character of the tidal field (\ref{tidal.V}), and hence, it cannot be employed to obtain the statistical average of the interaction energy $W_{AB}$. Denoting by $\mathcal{H}_{A}=H_{A}+W_{tidal}$ the full Hamiltonian of the system A, its entropy $S_{A}=k\ln \Omega_{A}$ can be estimated from the calculation of the density of states $\Omega_{A}$:
\begin{equation}\label{quasi-erg}
\Omega_{A}\left[E_{A},N|\varepsilon_{c}\right]=\frac{1}{N!}\int_{\mathcal{P}_{R}}\delta\left[E_{A}-\mathcal{H}_{A}\right]
\delta\left[\mathbf{P}_{A}\right]\delta\left[\mathbf{R}_{A}\right]\frac{d^{3N}\mathbf{r}d^{3N}\mathbf{p}}{(2\pi \hbar)^{3N}}.
\end{equation}
The integration over the $N$-body phase space $\mathcal{P}_{R}$ should be \emph{regularized} in similar fashion as the one considered by quasi-ergodic model \cite{Vel.QEM} to account for the energy threshold $\varepsilon_{c}$ for particles evaporation. This study is currently being developed in a separate work \cite{Gomez-Leyton2015}, which is a very important task to address some other related problems. In particular, this study sets the basis for a possible generalization of conventional canonical or gran canonical ensembles to astrophysical situations \cite{Morales}, as well as the derivation of a suitable \emph{Langevin-type equation} to describe the energy interchange between the astrophysical systems.

Let us attempt to obtain a roughly estimation of the statistical average of the quadrupole tensor component $Q^{xx}_{A}$:
\begin{equation}
Q^{xx}_{A}=\sum_{i}m_{i}\left[2x^{2}_{i}-y^{2}_{i}-z^{2}_{i}\right]
\end{equation}
by introducing some suitable simplifying assumptions. For example, let us assume that each astrophysical system is mainly located inside a certain inner region (e.g.: the called isothermal core of particles distribution) with average density $\rho_{A}$. The crudest description for dynamical effects of gravitational forces inside this region corresponds to the harmonic potential field:
\begin{equation}
W_{g}\simeq W_{0}+\sum_{i}\frac{1}{2}m_{i}\omega^{2}_{ff}\mathbf{r}^{2}_{i},
\end{equation}
where $\omega_{ff}$ is the \emph{free-fall frequency}:
\begin{equation}\label{free-fall.omega}
\omega^{2}_{ff}=4\pi G\rho_{A}/3.
\end{equation}
Note that its associated time period $\tau_{ff}=2\pi/\omega_{ff}$ represents the typical timescale of internal degrees of freedom. The total potential field $W$ can be expressed as:
\begin{equation}
W=W_{g}+W_{tidal}=W_{0}+\sum_{i}\frac{1}{2}m_{i}\left[\omega^{2}_{x}x^{2}_{i}+\omega^{2}_{y}y^{2}_{i}+
\omega^{2}_{z}z^{2}_{i}\right],
\end{equation}
where the frequencies $\left(\omega_{x},\omega_{y},\omega_{z}\right)$ are given by:
\begin{equation}
\omega^{2}_{x}=\omega^{2}_{ff}-\omega^{2}_{R}(1+2\xi_{A}),\mbox{ }\omega^{2}_{y}=\omega^{2}_{ff}-\omega^{2}_{R}(1-\xi_{A})\mbox{ and }\omega^{2}_{z}=\omega^{2}_{ff}+\omega^{2}_{R}\xi_{A}.
\end{equation}
Assuming the \emph{equipartition relations} (due to isothermal character of particles distributions in innermost region of each astrophysical system):
\begin{equation}
\left\langle\frac{1}{2}m_{i}\omega^{2}_{x}x^{2}_{i}\right\rangle=\left\langle\frac{1}{2}m_{i}\omega^{2}_{y}y^{2}_{i}\right\rangle=
\left\langle\frac{1}{2}m_{i}\omega^{2}_{z}z^{2}_{i}\right\rangle=\frac{1}{2}kT,
\end{equation}
one obtains the following estimation:
\begin{equation}
\left\langle Q^{xx}\right\rangle=N_{A}kT_{A}\frac{2\omega^{2}_{y}\omega^{2}_{z}-\omega^{2}_{x}\left(\omega^{2}_{y}+\omega^{2}_{z}\right)}
{\omega^{2}_{x}\omega^{2}_{y}\omega^{2}_{z}},
\end{equation}
where $N_{A}$ is the number of particles inside the inner region. Taking into consideration that the timescale of collective motions is much greater than the one of internal degrees of freedom, $\omega_{ff}\gg \omega_{R}$, one obtains the following expression:
\begin{equation}
\left\langle Q^{xx}\right\rangle\simeq \frac{9}{16\pi^{2}}\frac{N_{A}kT_{A}}{G^{2}\rho^{2}_{A}}\left(6\xi_{A}+1\right)
\omega^{2}_{R}\equiv\frac{1}{G}\left(6\xi_{A}+1\right)q_{A}\omega^{2}_{R},
\end{equation}
where the quantity $q_{A}$ was introduced:
\begin{equation}\label{qq}
q_{A}=\frac{9}{16\pi^{2}}\frac{N_{A}kT_{A}}{G\rho^{2}_{A}}
\end{equation}
that depends on the internal thermodynamical state of the system A. Using this last expression, the statistical average of the quadrupole term (\ref{quadrupole}) can be estimated as follows:
\begin{equation}\label{astro.thermal.contact}
\left\langle W_{AB}\right\rangle\simeq-\frac{GM_{A}M_{B}}{2D}-
\frac{G(M_{A}+M_{B})^{2}}{2D^{6}}\left[\xi_{A}(6\xi_{A}+1)q_{A}+\xi_{B}(6\xi_{B}+1)q_{B}\right].
\end{equation}
Here, it was employed the relation $\mathbf{P}^{2}/2\mu_{AB}=GM_{A}M_{B}/2D$ for kinetic and gravitational energy for a motion along a circular orbit. If the separation distance $D$ between the astrophysical systems is sufficiently large, the second term in the estimation (\ref{astro.thermal.contact}) represents a small contribution that introduces a perturbation for the collective motions. Thermodynamically, the quadrupole-orbit coupling in (\ref{quadrupole}) represents the \emph{astrophysical counterpart of thermal contact} for the collective motions. For small distances, the quadrupole contribution in (\ref{astro.thermal.contact}) grows faster than the Newtonian contribution, which could lead to the breakdown of effective separability between the astrophysical systems, and their eventual merge.

Quadrupole terms in (\ref{quadrupole}) have a significant influence on the thermodynamical behavior of each system. In fact, these terms determine the stable branch between the point of gravitational collapse $u_{A}$ and up to the point of evaporative disruption $u_{C}$ shown in figure \ref{King.eps}. This contribution provides a fairly estimation for the tidal radius $R_{t}$ of globular clusters \cite{Binney}:
\begin{equation}\label{tidal.quad}
R_{A}\equiv R_{t}\simeq D\left(M_{A}/2M_{B}\right)^{1/3}
\end{equation}
for the asymptotic case $M_{A}\ll M_{B}$. This radius characterizes the distance where the gravitational influence of system A is overcome by gravitational field of system B, and hence, it defines both the linear size of system A and its threshold energy $\varepsilon_{c}$ for particles escape. For a better understanding, let us consider the estimation (\ref{tidal.quad}) and neglect the quadrupole deformation of the system B, $q_{B}\simeq0$. Taking into account that $N_{A}kT_{A}\sim GM^{2}_{A}/R_{A}$, one obtains $q_{A}\sim R^{5}_{A}$, so that, the quadrupole contribution in (\ref{astro.thermal.contact}) is of order $(GM_{A}M_{B}/D) \times (M_{A}/M_{B})^{2/3}\ll GM_{A}M_{B}/D$. The energy contribution of quadrupole-orbit coupling is much smaller than the Newtonian energy contribution of collective motions. However, this energy is fully comparable to the internal energy of the system A, $\left|U_{A}\right|\sim GM^{2}_{A}/R_{A}\sim (GM_{A}M_{B}/D)\times \left(M_{A}/M_{B}\right)^{2/3}\ll GM_{A}M_{B}/D$. These roughly estimations evidence that the factor $\phi_{A}$ defined in  (\ref{factors.coll}) is not necessarily small.

\subsection{On the role of quadrupole-orbit coupling and angular momentum on the stability of a binary system}\label{stab.coll}

\begin{figure}
\centering
\begin{minipage}{.5\textwidth}
  \centering
  \includegraphics[width=3.0in]{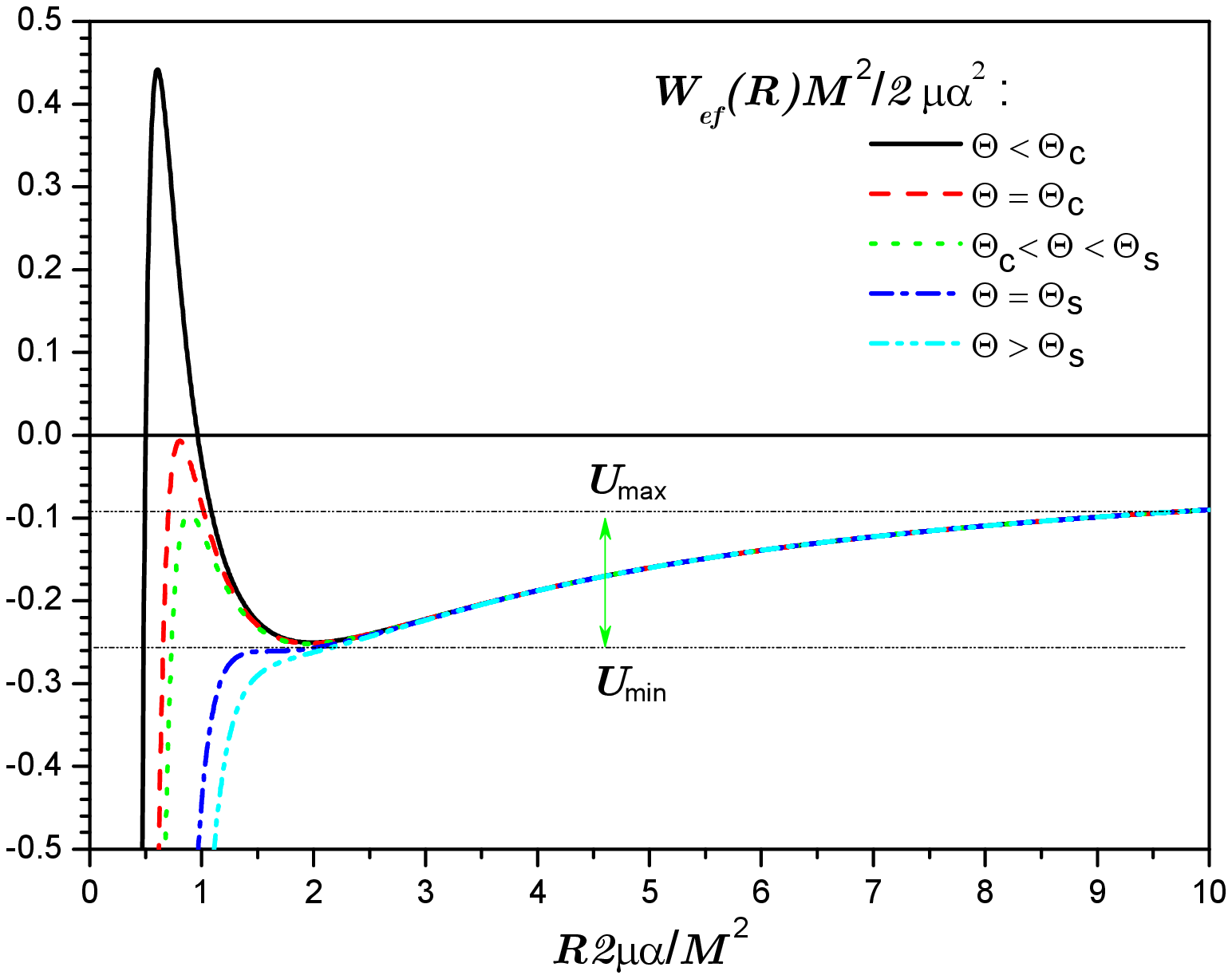}
\end{minipage}%
\begin{minipage}{.5\textwidth}
  \centering
  \includegraphics[width=2.8in]{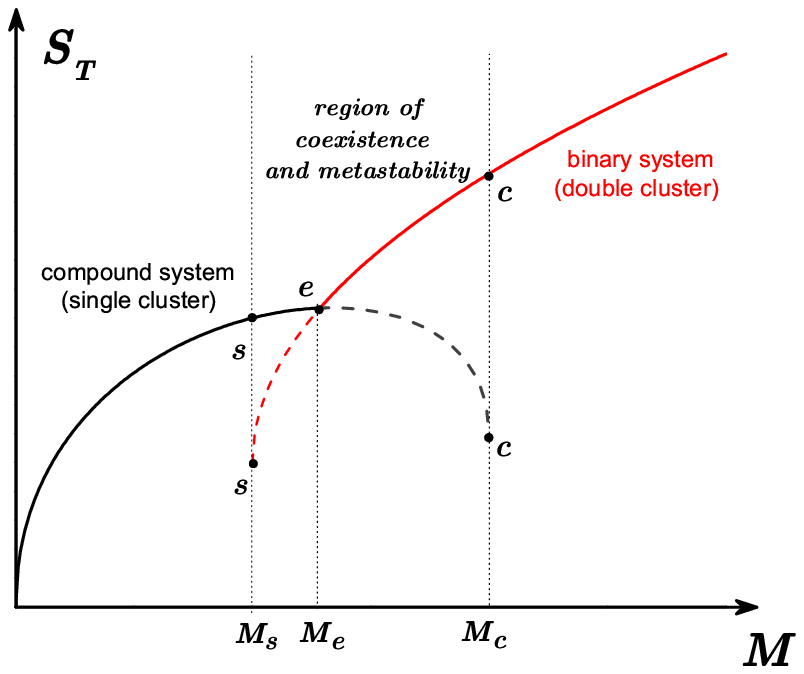}
\end{minipage}
\caption{Left panel: Radial dependence of the effective potential (\ref{eff.V}) for different values of quadrupole parameter $\Theta$. Vertical and horizontal scales in this plot are expressed in units derived from the total angular momentum $\mathbf{M}$. Right panel: Schematic representation of the dependence of the total entropy $S_{T}$ on the angular momentum. Two branches for low and high values of momentum, where there is an intermediate region with a coexistence of single and double cluster configurations.}\label{momentum.eps}
\end{figure}

As already shown, the quadrupole terms in (\ref{quadrupole}) introduce, in average, a contribution $1/R^{6}$ in the gravitational interaction between the astrophysical systems. This type of radial dependence should be also observed for collective motions with non-circular orbits, so that, the dynamics of collective motions could be described by the effective Hamiltonian:
\begin{equation}
H_{C}=\mathbf{P}^{2}/2\mu+W(\mathbf{R}).
\end{equation}
The total potential $V(\mathbf{R})$ is given by:
\begin{equation}\label{VV}
W(\mathbf{R})=-\alpha/\left|\mathbf{R}\right|-\alpha\Theta/\left|\mathbf{R}\right|^{6}
\end{equation}
with parameters:
\begin{equation}
\mu=\frac{M_{A}M_{B}}{M_{A}+M_{B}},\:\alpha=GM_{A}M_{B}\mbox{ and }\Theta=\frac{\xi_{A}(6\xi_{A}+1)q_{A}+\xi_{B}(6\xi_{B}+1)q_{B}}{2\xi_{A}\xi_{B}}.
\end{equation}
The second term of potential (\ref{VV}) is fully analogous to the leading term of Van der Waals potential among non-polar molecules and atoms \cite{Reichl}:
\begin{equation}
W_{VdW}(R)=-\epsilon\left[\left(\frac{\sigma}{R}\right)^{6}-\left(\frac{\sigma}{R}\right)^{12}\right].
\end{equation}
The same one becomes dominant if the systems A and B are sufficiently close each other. Accordingly, this term can explain an eventual the merge of these systems. Due to non-integrability of this type of potential \cite{Broucke}, the presence of this term could lead in certain conditions to the dynamical instability of collective motions, that is, the existence of a chaotic behavior. The conservation of the total angular momentum $\mathbf{M}$ for this situation enables the study of collective motions in the orbital plane. Considering polar coordinates, the square of the linear momentum can be spread into the radial and the angular contribution as $\mathbf{P}^{2}=\mathbf{P}^{2}_{R}+\mathbf{M}^{2}/R^{2}$. This procedure enables the introduction of the effective potential for the radial evolution as follows \cite{Goldstein}:
\begin{equation}\label{eff.V}
W_{ef}(R)=\mathbf{M}^{2}/2\mu R^{2}-\alpha/R-\alpha\Theta/R^{6}.
\end{equation}
As clearly shown in the left panel of figure \ref{momentum.eps}, this effective potential provides valuable information about the role of total angular momentum $\mathbf{M}$ and the quadrupole-orbit coupling on the stability of this binary astrophysical system. For a given value $M=\left|\mathbf{M}\right|$ of the modulus of total angular momentum $\mathbf{M}$, there exist two critical values for quadrupole parameter $\Theta$,  $\Theta_{c}=\left(2M^{2}/5\mu\alpha\right)^{5}/4$ and $\Theta_{s}=\left(4M^{2}/5\mu\alpha\right)^{5}/24$. The first value $\Theta_{c}$ guarantees that the local maximum of effective potential (\ref{eff.V}) is zero. The second value $\Theta_{s}$ guarantees that the only stationary point of potential (\ref{eff.V}) is an inflexion point. Denoting by $U_{C}<0$ the energy of bounded collective motions, the binary astrophysical system is always stable when $\Theta\leq\Theta_{c}$. For the interval $\Theta_{c}<\Theta<\Theta_{s}$, stable collective motions are only possible for a certain energy window $U_{min}\leq U_{C}<U_{max}<0$. The minimal value $U_{min}$ (circular orbits) weakly depends on the quadrupole parameter $\Theta$, and its value is close to $U_{min}\simeq -0.5\alpha^{2}\mu/M^{2}$. Within the energy window $U_{max}<U_{C}<0$, collective motions are unstable and the binary system collapses to form a single structure. Finally, the collective motions are always unstable when $\Theta\geq \Theta_{s}$. The present picture can be rephrased in terms of the angular momentum for a fixed value of quadrupole parameter $\Theta$. In this case, the critical values of quadrupole parameter provides two critical values for the modulus of the angular momentum $M_{c}$ and $M_{s}$:
\begin{equation}
M_{c}=\left[\left(4\Theta\right)^{1/5}5\mu\alpha/2\right]^{1/2}\mbox{ and }M_{s}=\left[\left(24\Theta\right)^{1/5}5\mu\alpha/4\right]^{1/2}.
\end{equation}
Accordingly, binary system is stable if the modulus of angular momentum $M>M_{c}$. The merge of these systems is always produced if the modulus of angular momentum $M<M_{s}$. Finally, both configurations are possible for intermediate values of the angular momentum $M_{s}<M<M_{c}$.

The estimates (\ref{qq}) for the functions $(q_{A},q_{B})$ only provide a qualitative dependence on the internal thermodynamic conditions of each system. Their values decrease with the growth of the inner density of each astrophysical system. Accordingly, the collective motions gain stability when the interacting astrophysical systems exhibit more compact internal structures, that is, particles distributions with more dense cores. The most compact structures are always observed for low energies, which also correspond to configurations with negative heat capacities (see in figure \ref{King.eps}). Although in an indirect way, the coexistence of two interacting astrophysical systems with negative heat capacity is convenient for the stability of collective motions. In summary, the stability binary astrophysical system should be possible for low energies and large values of the angular momentum. The merge of two astrophysical systems into a single astrophysical structure should be observed for low energies and low values of the angular momentum. Finally, both type of configurations should observed for intermediate values of angular momentum.

A more complete picture about dynamical and thermodynamical stability for the present situation is obtained by including the incidence of thermal fluctuations. Formally, this goal can be achieved in the framework of classical fluctuation theory \cite{Reichl} by considering concrete expressions for the entropies of each system $S_{A}$ and $S_{B}$ and the energetic constraint (\ref{energ.constraint}). In principle, one can expect that dependence of total entropy $S_{T}=S_{A}+S_{B}$ on the angular momentum $\mathbf{M}$ exhibits a behavior as the one schematically represented in figure \ref{momentum.eps}. Here, I have included the sketch of the stable branch that corresponds to the compound system resulting from the merge of the astrophysical systems. Notice the existence of a region with intermediate values of angular momentum $M_{s}<M<M_{c}$ where one observes the coexistence of both configurations. It is clear that for values of $M$ near but below the point $M_{c}$, the binary system is not the only possible configuration, but it should be more probable than the compound system. Alternatively, the same qualitative behavior should be expected for values of $M$ close but above the point $M_{c}$ for the configurations of compound system. Therefore, these stable branches will intersect in a certain value of the angular momentum $M_{e}$. Configurations of binary system within the interval $M_{s}<M<M_{e}$ (dashed red line) are \emph{metastable states} since the configurations of compound system are more probable (exhibit a greater entropy). Alternatively, configurations of compound system within the interval $M_{e}<M<M_{c}$ (dashed black line) are also metastable states.

\begin{figure}
\begin{flushright}
\includegraphics[width=4.5in]{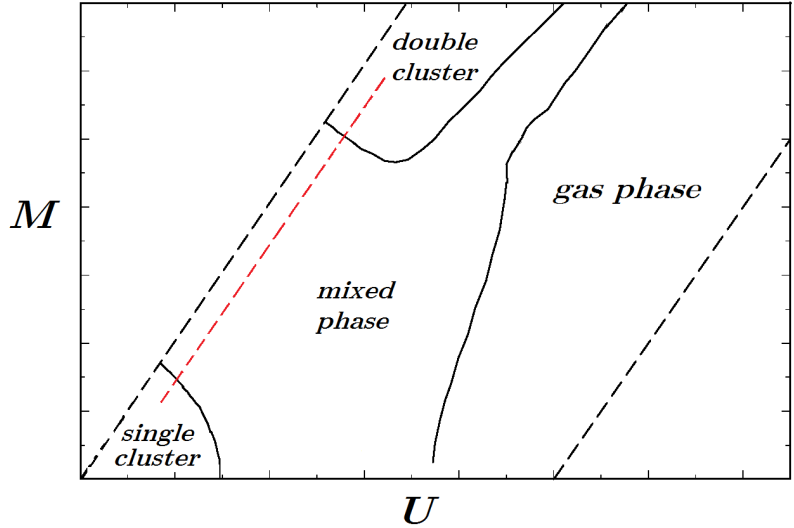}
\end{flushright}
\caption{Phase diagram reported by Votyakov and co-workers in Ref.\cite{Votyakov}. Dependence of the total entropy along the dashed red line would describe qualitative behavior shown in left panel of figure \ref{momentum.eps}.}\label{phased.eps}
\end{figure}

The picture described above agrees well with results reported by Votyakov and co-workers \cite{Votyakov}. These authors considered an astrophysical system of identical non-relativistic particles with hard-cores, which was enclosed into a container to avoid evaporation. They analyzed the thermodynamic behavior of this system in the microcanonical description by imposing the conservation of energy and the total angular momentum. The phase diagram of this model is reproduced in figure \ref{phased.eps}. Configurations with large positive energies are now accessible, which correspond to a gas phase as the case of Antonov isothermal model. Configurations with very low energies exhibit particles distributions with a compact core, whose density is determined by the hard core-radius $a$ of the particles as $n_{c}\sim 3/4\pi a^{3}$. Such configurations exhibit positive heat capacities, and they correspond to the post-collapse phases. The thermodynamic behavior in this region crucially depends on the total angular momentum. While one observes a single astrophysical structure (a singe cluster) for low values of the total angular momentum, the system spontaneously develops binary configurations (a double cluster) for values of angular momentum sufficiently large. These phases are separate by a region of phase coexistence (or mixed phases), where the closed system exhibits both single and double clusters configurations for intermediate values of angular momentum and \emph{negative values of the heat capacity}. In principle, the qualitative behavior of the total entropy  $S_{T}$ represented in right panel of figure \ref{momentum.eps} should be observed along configurations on the red dashed line highlighted in this phase diagram.

\subsection{Estimation of entropy using virial relations and associated thermodynamic limit}\label{thermo.L.sect}

Taking into consideration Gibbs derivation of canonical ensemble, the possible extension of this analysis for astrophysical situations will demand to invoke a suitable thermodynamic limit for the system that acts as \emph{surrounding} (see in figure \ref{CoupledBaths.eps}). Of course, thermodynamic limit of conventional extensive systems (\ref{ext.TL}) is nonapplicable. In fact, astrophysical systems do not follow a universal behavior as the case of extensive systems. Nevertheless, one could obtain some analogous scaling relations among physical observables in certain limit situations. Scaling relations depend on the concrete features of each astrophysical model, as example, if they are composed of point particles or hard-core spheres, or the relativistic or non-relativistic character of their microscopic dynamics, etc. In fact, the proper limit $N\rightarrow\infty$ is meaningless for the case of compact objects in astrophysics (e.g.: white dwarfs, neutron stars, etc.) because of the relativistic character of these astrophysical structures imposes certain upper limits to the total mass to guarantee their stability \cite{Chandrasekhar.limit,Oppenheimer,Tolman}.

It is possible to introduce a thermodynamic limit for the self-gravitating gas of identical non-relativistic point particles. All models previously discussed in this work belong to this type of astrophysical systems. Without any other external influence, this astrophysical system does not exhibit any characteristic linear dimension or characteristic energy (in the framework of classical mechanics), and the same one suffers from the incidence of evaporation. This situation was previously considered during the analysis heat capacity of astrophysical systems using the virial theorem, subsection \ref{Eddington}. Let us now apply the same reasonings to obtain a roughly estimation for the microcanonical entropy for this astrophysical system.
Considering the virial relations for the total kinetic energy $K$ and the total potential energy $W$:
\begin{equation}
\left\langle K\right\rangle=N\left\langle\mathbf{p}^{2}/2m\right\rangle\sim \vert U\vert\mbox{ and }\vert U\vert \sim-\left\langle W\right\rangle /2\sim GM^{2}/R.
\end{equation}
one obtains the following characteristic momentum $\vert \mathbf{p}_{c}\vert \sim\left(2m \vert U\vert /N\right)^{1/2}$ and radius $R_{c}\sim GM^{2}/\vert U\vert $. These characteristic quantities can be employed to estimate the accessible phase space volume $\Omega$ (number of micro-states):
\begin{equation}
\Omega\sim \left(p_{c}R_{c}/\hbar\right)^{3N}/N!,
\end{equation}
which yields the following estimation for the system entropy $S=k\ln \Omega$:
\begin{equation}\label{mic.entropy}
S\sim
Nk\ln\left[\left(\frac{G^{2}m^{5}}{2\hbar^{2}}\frac{N^{3}}{\left\vert U\right\vert }\right)^{\frac{3}{2}}\frac{e}{N}\right].
\end{equation}
As usual, the factorial factor $N!$ was introduced to take into account the particles identity. It is noteworthy that this situation is related to the one described by Plummer model. In fact, the estimation (\ref{mic.entropy}) can be improved starting from the usual definition:
\begin{equation}
S=-k\int f(\mathbf{r},\mathbf{p})\ln \left[f(\mathbf{r},\mathbf{p})\right]\frac{d^{3}\mathbf{r}d^{3}\mathbf{p}}{(2\pi \hbar)^{3}}
\end{equation}
and considering the quasi-stationary one-body distribution (\ref{Plumer.model}). However, this procedure can hardly modify some numerical constants in the argument of logarithmic function. The existing dependence on energy $U$ and physical constants remains unchanged since the logarithm argument is mostly determined by the normalization constant $A$ described in equation (\ref{AA.Plummer}).

Let us analyze some consequences of the previous result. Considering that the total energy $U$ is negative, that is $\left|U\right|\equiv-U$, the virial estimation of heat capacity $C=-3Nk/2$ is re-obtained by combining the estimation (\ref{mic.entropy}) and definition (\ref{mic.def}). According to estimation (\ref{mic.entropy}), the entropy per particle $S/N$ remains finite in the thermodynamic limit $N\rightarrow\infty$ considering the following non-extensive scalings for the total energy and the characteristic radius $R$:
\begin{equation}\label{non.extensive.th}
U/N^{7/3} = const \mbox{ and } RN^{1/3} = const.
\end{equation}
The present scaling relations have been obtained for more sophisticated astrophysical models of non-relativistic point particles \cite{Vel.QEM,Chavanis2002,Chavanis2005}. This thermodynamic limit evidences that astrophysical systems are non-extensive. One could find strange that the linear size $R$ of the astrophysical system decreases with the growth of the number of particles $N$. However, there is a simple explanation: the size $R$ is reduced because of the system gravitational field becomes stronger with the grow of its total mass $M$.

\subsection{Other thermodynamic limits proposed in the literature}\label{TL.others}

Recently, other investigators have claimed the physical relevance of other different thermodynamics limits for this same type astrophysical systems (with identical non-relativistic point particles). A particular example is the called \emph{diluted-limit}:
\begin{equation}\label{diluted}
U/N = const \mbox{ and } R/N = const
\end{equation}
introduced by de Vega and Sanchez \cite{Vega1}, where the energy $U$ appears as an extensive quantity. This thermodynamic limit exemplifies the \emph{Kac prescription} \cite{Kac}, where some independent parameters of a given model (e.g.: the linear dimension of container $R$) are re-scaled with the number of particles $N$ in order to force the extensivity of the total energy $U$, and hence, the intensive character of the temperature $T$. Other proposal is the thermodynamic limit considered by Cerruti-Sola et al \cite{Cerruti}:
\begin{equation}\label{cerruti.tl}
U/N^{5/3} = const \mbox{ and } R/N^{1/3} = const,
\end{equation}
which now guarantees the extensivity of the system volume $V\sim R^{3}\propto N$. These last authors presented a nice demonstration about a data collapsing of microcanonical caloric curves for different number of particles $N$ once the above scaling laws for energy are introduced.

\emph{What criterium of thermodynamic limit is the fundamental one?} In general, the scaling behaviors of macroscopic observables with $N$ are irrelevant whenever the state equations of a given astrophysical system are expressed in terms of dimensionless variables like $u=UR/GM^{2}$ and $t=kTR/GMm$. This criterium was adopted in figures \ref{Antonov.eps} and \ref{King.eps} to describe the microcanonical caloric curves of Antonov isothermal model and King model. However, I think that the scaling laws should not be so arbitrary. If they exist in certain limit situations, it is natural to expect that they would be \emph{intrinsic properties} for a given thermodynamic system. In particular, this question could be clarified through the connections of these scaling laws with other macroscopic behaviors, such as dynamic relaxation towards equilibrium, the correspondence between quantum and classical (relativistic and non-relativistic) descriptions, etc.

Let us illustrate this argument considering that the astrophysical system of point particles is a cloud of hydrogen atoms with mass $m=m_{p}$. According to the estimation (\ref{mic.entropy}), its total non-relativistic energy is of order $U\sim-G^{2}m^{5}_{p}/\hbar^{2}N^{7/3}$. The licitness of this estimation requires that the absolute value of non-relativistic total energy $U$ should be lower than the system rest energy $U_{0}=Nm_{p}c^{2}$, which is fulfilled when the total mass of the system $M=Nm_{p}$ obeys the following upper bound $M_{c}$:
\begin{equation}\label{Mc}
M\leq M_{c}=\left(\frac{\hbar c}{G}\right)^{\frac{3}{2}}\frac{1}{m^{2}_{p}}\simeq 29.2M_{\odot}.
\end{equation}
The quantity $M_{c}$ is the characteristic mass that enters in stability limits of stars \cite{Chandrasekhar.limit,Oppenheimer,Tolman}. As already commented by Chandrasekhar \cite{Chandrasekhar}, in the same way that the Bohr radius $a_{B}=\hbar^{2}/e^{2}m_{e}$ explains the properties of atoms, the characteristic constant (\ref{Mc}) explains why the stars are as they are. This result can be obtained from the non-extensive thermodynamic limit (\ref{non.extensive.th}) only. In fact, this scaling laws arises for self-gravitating non-relativistic degenerate gas of fermions in white dwarfs models. For example, the radius $R$ of a white dwarf obeys the following scaling relation with the total mass $M$:
\begin{equation}\label{dwarfs.radius}
R\propto \frac{\hbar^{2}\beta^{5/3}}{G m_{e}m^{5/3}_{p}}M^{-1/3}
\end{equation}
with $\beta=\bar{Z}/\bar{A}$ is the average number of electrons per nucleon, while $m_{e}$ is the electron mass. This is the same scaling law of the linear size $R$ described in equation (\ref{non.extensive.th}).

\subsection{Astrophysical counterpart of Gibbs-Duhem relation}\label{sub.Gibbs-Duhem}

Let us reconsider the expression of entropy derived by de Vega and Sanchez:
\begin{equation}\label{SS.Vega}
S(U,L,N)=k\ln\left[\frac{N^{3N-2}m^{9N/2-2}L^{3N/2+1}G^{3N/2-1}}{N!\Gamma\left(\frac{3N}{2}\right)\left(2\pi \hbar^{2}\right)^{3N/2}}\right]+k\ln w(u,N)
\end{equation}
in the framework of Antonov isothermal model \cite{Vega1}. For the sake of convenience, I have restituted here Planck constant $\hbar$, which was previously omitted by these authors by setting its value as $\hbar\equiv 1$. The function $w(u,N)$ is given by the multidimensional integral:
\begin{equation}\label{wVega}
w(u,N)=\int \prod_{i}d^{3}\mathbf{x}_{i}\left[u+\upsilon_{N}\left(\mathbf{x}\right)\right]^{3N/2-1}
\Theta\left[u+\upsilon_{N}\left(\mathbf{x}\right)\right],
\end{equation}
where $\upsilon_{N}\left(\mathbf{x}\right)$ defined as:
\begin{equation}
\upsilon_{N}\left(\mathbf{x}\right)=
\frac{1}{N^{2}}\sum_{i<j}\frac{1}{\left|\mathbf{x}_{i}-\mathbf{x}_{j}\right|},
\end{equation}
with $u=UL/GM^{2}$ being the dimensionless energy variable. Rigorously, equation (\ref{wVega}) is a divergent integral. The same one can be partially \emph{regularized} in the framework of mean field (continuum) approximation. Other stronger regularization schemes are possible, such as the use of a small cutoff radius $a$ to avoid divergence of Newtonian potential:
\begin{equation}
\varphi_{a}(\mathbf{x}_{i},\mathbf{x}_{j})=\left\{
\begin{array}{cc}
  1/a,& \mbox{ if }\left|\mathbf{x}_{i}-\mathbf{x}_{j}\right|<a ,\\
  1/\left|\mathbf{x}_{i}-\mathbf{x}_{j}\right|, & \mbox{ otherwise}.
\end{array}
\right.
\end{equation}
The characteristic linear dimension $L$ of the container was previously employed as the length unit in order to express the integral expression (\ref{wVega}) into dimensionless form. The particles positions $\mathbf{r}$ were rewritten into dimensionless form as $\mathbf{r}_{i}\rightarrow\mathbf{x}_{i}=\mathbf{r}_{i}/L$.

Let us pay attention to the first term of the entropy (\ref{SS.Vega}). Considering Stirling approximation for factorial $\ln n!\simeq n\ln n-n$ (and Gamma function), the leading order of this terms is given by:
\begin{equation}\label{additional.term}
\sim\frac{3}{2}Nk\ln\left[\frac{e^{5/3}LN^{1/3}}{3\pi\ell_{G}}\right],
\end{equation}
where $\ell_{G}=\hbar^{2}/Gm^{3}$ is a characteristic length with a certain quantum significance. Formally, this is a gravitational counterpart of Bohr radius $r_{B}=\hbar^{2}/e^{2}m_{e}$. The physical meaning of the characteristic length $\ell_{G}$ can be understood by analyzing the applicability of classical description of the present self-gravitating gas. Denoting by $d\sim n^{-1/3}$ the characteristic separation among the particles for the gas with mean density $n$, the quantum uncertainty of the total kinetic energy $K$ must satisfy the following condition:
\begin{equation}
N\frac{1}{2m}\left(\frac{\hbar}{d}\right)^{2}\ll \left\langle K\right\rangle\sim -\frac{1}{2}\left\langle W\right\rangle\sim \frac{GM^{2}}{2L}.
\end{equation}
Taking into account the expressions $M=Nm$ and $n\sim N/L^{3}$, one obtains:
\begin{equation}\label{classical.limit}
LN^{1/3}\gg \ell_{G}.
\end{equation}
Accordingly, the product $LN^{1/3}$ should remain constant in the thermodynamic limit $N\rightarrow +\infty$ to guarantee the licitness of classical description, equation (\ref{classical.limit}). If quantum behavior turns relevant, the previous result can be reinterpreted to provide a fairly estimation for the system linear size as a function on its total mass $M$:
\begin{equation}
L\propto \frac{\hbar^{2}}{Gm^{8/3}}M^{-1/3}.
\end{equation}
Considering $m$ as the neutron mass, this last expression is simply the non-relativistic expression for the radius of a neutron star. This result is fully analogous to expression (\ref{dwarfs.radius}) for the typical radius of white dwarfs.

It possible to verify that the second term of expression (\ref{SS.Vega}) exhibits an extensive behavior with the number of particles $N$:
\begin{equation}\label{leading.term}
\lim_{N\rightarrow+\infty}\frac{1}{N}k\ln w(u,N)=s_{R}(u),
\end{equation}
where the \emph{reduced entropy} $s_{R}(u)$ is a function on the dimensionless energy $u$ only. Considering the thermodynamic relations:
\begin{equation}\label{thermo.identity}
\frac{1}{T}=\frac{\partial }{\partial U}S(U,L,N)\mbox{ and }\frac{1}{T}p=\frac{\partial }{\partial V}S(U,L,N),
\end{equation}
where $V\propto L^{3}$ is the volume of the system, one obtains the state equation:
\begin{equation}\label{state.equation}
3pV=\frac{3}{2}NkT+U\equiv 2\left\langle K\right\rangle+\left\langle W\right\rangle.
\end{equation}
This last relation drops to the virial theorem (\ref{virial}) once the equipartition relation $\left\langle K\right\rangle=3NkT/2$ is considered. The leading order term (\ref{additional.term}) also contributes during calculation of the second thermodynamic identity in (\ref{thermo.identity}). Additionally, the \emph{chemical potential} $\mu$ is obtained from the identity:
\begin{equation}
-\frac{1}{T}\mu=\frac{\partial}{\partial N}S(U,L,N),
\end{equation}
which yields:
\begin{equation}
-\frac{1}{T}\mu N=S(U,L,N)+\frac{1}{2}Nk-2\frac{1}{T}\left\langle U\right\rangle.
\end{equation}
Rephrasing the entropy $S(U,L,N)$ in terms of the volume $V\propto L^{3}$ as $S(U,V,N)$, as well as the state equation (\ref{state.equation}) as follows:
\begin{equation}
\frac{1}{2}Nk=\frac{pV}{T}-\frac{1}{3}\frac{U}{T},
\end{equation}
one arrives at a sort of \emph{Gibbs-Duhem relation}:
\begin{equation}\label{Duhem-Gibbs}
S(U,V,N)=\frac{1}{T}\left[\frac{7}{3}U-pV-\mu N\right].
\end{equation}
Note that the previous result is straightforwardly obtained from the scaling behavior:
\begin{equation}\label{scaling.sgg}
S(\alpha^{7/3}U,\alpha^{-1}V,\alpha N)=\alpha S(U,V,N),
\end{equation}
which is fully consistent with the thermodynamic limit (\ref{non.extensive.th}). By itself, this last result provides a very strong support about the relevance of thermodynamic limit (\ref{non.extensive.th}) over any other proposal.

\subsection{Relaxation time and thermodynamic limit}

Let us analyze the relation between the thermodynamic limit (\ref{non.extensive.th}) and the relaxation time of gravitational systems. According to estimations derived from Chandrasekhar theory of dynamical friction \cite{Chandrasekhar}, the collisional relaxation time $\tau_{cr}$ can be expressed as follows:
\begin{equation}\label{collisional.relax}
\tau_{cr}=\frac{N}{\ln N}\tau_{d},
\end{equation}
where $\tau_{d}=1/(G\rho)^{1/2}$ is the characteristic time of microscopic dynamics (related to free-fall frequency of particles (\ref{free-fall.omega}) in a self-gravitating gas with average mass density $\rho$). Taking into consideration that characteristic density $\rho$ scales with the number of particles $N$ as $\rho\sim N/R^{3}\sim N^{2}$, one obtains that the collisional relaxation time scales as $\tau_{cr}=O(1/\ln N)$. It is noteworthy that the alternative thermodynamic limits (\ref{diluted}) and (\ref{cerruti.tl}) lead to the divergence of relaxation time when $N\rightarrow+\infty$ as $\tau_{cr}=O(N^{2}/\ln N)$ and $\tau_{cr}=O(N/\ln N)$, respectively. A divergence of relaxation time would imply the \emph{non-commutativity} of thermodynamic limit and the infinite time limit necessary for convergence of temporal expectation values:
\begin{equation}\label{non-commutativity}
\lim_{N\rightarrow+\infty}\lim_{t\rightarrow+\infty}O(t,N)\neq\lim_{t\rightarrow+\infty}\lim_{N\rightarrow+\infty}O(t,N).
\end{equation}
The use of these alternative thermodynamic limits implies an ill-defined $N$-dependence of the system temporal evolution. Among the three proposals considered in this work, the thermodynamic limit (\ref{non.extensive.th}) is the only one that fulfils commutativity of these two infinite limits. Its non consideration should lead to dynamical anomalies in numerical simulations \cite{Taruya}.

The thermodynamic limit (\ref{non.extensive.th}) \emph{almost} guarantees that the collisional relaxation time (\ref{collisional.relax}) becomes a $N$-independent quantity. This observation suggests the existence of a certain connection between the thermodynamic limit that guarantees the extensivity of entropy and the timescale of the system macroscopic relaxation. If exist this connection, the relevant relaxation timescale $\tau_{r}$ for astrophysical systems should exhibit the form:
\begin{equation}\label{tau.R}
\tau_{r}\propto N\tau_{d}.
\end{equation}
Derivation of collisional relaxation time (\ref{collisional.relax}) involves a series of assumptions with a restricted applicability. For example, the appearance of the term $\ln N$ in the estimation (\ref{collisional.relax}) depends on the collective effects that should be considered to avoid the short-range and long-range divergences associated with Newtonian gravitation \cite{Chavanis2013}. Collisional kinetic equations, such as Vlasov-Landau and Boltzmann equations, can be derived from Liuville equation and its associated BBGKY hierarchy. However, such a procedure provides an inadequate treatment to the formation of bound configurations with two or more particles, which is a very important behavior in presence of gravitation.

\subsection{Other antecedents and connections}

The non-extensive thermodynamic limit (\ref{non.extensive.th}) is not exclusive of astrophysical models analyzed in this work. Its associated scaling laws (\ref{scaling.sgg}) also arise in framework of Thomas-Fermy theory for the asymptotic behaviors of atomic energy $U(Z)$ and density $\rho^{Z}(\mathbf{r})$ for an atom of charge $Z$ sufficiently large \cite{Thomas,Scott,Lieb1,Lieb2}:
\begin{equation}
U(Z)=C_{TF}Z^{7/3}+O(Z^{2})\mbox{ and }\rho^{Z}(\mathbf{r})\sim\rho^{Z}_{TF}(\mathbf{r})=Z^{2}\rho^{1}_{TF}\left(Z^{1/3}\mathbf{r}\right).
\end{equation}
This precedent result strongly reinforces the idea that the scaling behavior (\ref{scaling.sgg}) also apply for a three-dimensional system of non-relativistic point particles that interact among them by means of $1/r$  potential without mattering about their quantum or classical behavior. As already commented in subsection \ref{TL.others}, the thermodynamic limit (\ref{non.extensive.th}) is also relevant in Chandrasekhar theory of non-relativistic white dwarfs \cite{Chandrasekhar.limit}. This is a quite expected result because of this physical theory is a gravitational analogous to Thomas-Fermy theory of electrons in atoms \cite{Thomas}. In principle, the non-extensive thermodynamic limit (\ref{non.extensive.th}) should apply for pure non-neutral plasmas that are confined by external magnetic and electric fields, e.g.: a pure electron plasma \cite{Dubin}.

The generalization of Gibbs-Duhem relation for systems with long-range interactions was recently addressed by Latella and co-workers in Refs. \cite{Latella.2013,Latella.2015}. Roughly speaking, these authors shown that the non-extensive (or \emph{non-additive} in their terminology) properties certain systems will introduce an additional term $\mathcal{E}$ in the original Gibbs-Duhem relation as follows:
\begin{equation}\label{GD.Latella}
\mathcal{E}=U-TS+pV-\mu N,
\end{equation}
which they referred to as the \emph{replica energy}. They also showed that this quantity can be expressed in terms of the total potential energy $W$ as follows:
\begin{equation}
\mathcal{E}=-\sigma W\mbox{ with }\sigma=(d-\nu)/d
\end{equation}
for the particular case of a $d$-dimensional classical self-interacting gas of non-relativistic point particles enclosed inside an impenetrable container of linear dimension $R$:
\begin{equation}\label{generic.HH}
H_{N}(\mathbf{r},\mathbf{p})=\sum_{i}\frac{1}{2m}\mathbf{p}^{2}_{i}+\sum_{i<j}\frac{A_{ij}}{\left|\mathbf{r}_{i}-\mathbf{r}_{j}\right|^{\nu}}.
\end{equation}
If the exponent of radial dependence of interaction potential is $\nu=1$, the generic Hamiltonian system (\ref{generic.HH}) includes the self-gravitating gas with Newtonian gravitation for coupling coefficients $A_{ij}=-Gm_{i}m_{j}$ and the non-neutral Coulombian plasma when $A_{ij}=q_{i}q_{j}$.

In general, I am in fully agreement with all technical and conceptual aspects discussed by these authors. However, is formula (\ref{GD.Latella}) an appropriate extension of Gibbs-Duhem relation? Without specifying the concrete mathematical expression of the quantity $\mathcal{E}$, this expression seems to be more a definition for the replica energy $\mathcal{E}$ rather than a thermo-statistical identity. In principle, it is not necessarily a limitation. As a definition, this expression will be always applicable to any thermodynamic situation because of $\mathcal{E}$ \emph{is a measure about the deviation from extensive properties}. This means that replica energy, in a certain sense, is a relative notion. For example, one can invoke the same arguments to introduce an \emph{astrophysical counterpart} for the replica energy $\mathcal{E}_{a}$ as follows:
\begin{equation}
\mathcal{E}_{a}=\frac{7}{3}U-TS-pV-\mu N,
\end{equation}
which characterizes the deviation of the scaling properties of a given system with respect to the ones associated with a 3D astrophysical system composed of non-relativistic point particles. For the generic Hamiltonian system (\ref{generic.HH}), one obtains the following relation:
\begin{equation}
\mathcal{E}_{a}\equiv \frac{4}{3d}\left(d-3\right)K+\frac{1}{3d}\left(d-3\nu\right)W,
\end{equation}
where $K$ is the total kinetic energy. Notice that this quantity vanishes for the astrophysical case where $d=3$ and $\nu=1$.

Conventionally, Gibbs-Duhem relation is a thermodynamical identity that follows as a consequence of extensive properties of large short-range interacting systems:
\begin{equation}
S(\alpha U,\alpha N, \alpha V)=\alpha S(U,N,V)\Rightarrow S=\frac{1}{T}\left(U+pV-\mu N\right).
\end{equation}
It is easy to realize that the astrophysical relations (\ref{Duhem-Gibbs}) and (\ref{scaling.sgg}) exhibit the same geometric significance. Latella and co-workers do not enter to analyze the scaling laws and the thermodynamic limit associated with the generic Hamiltonian system (\ref{generic.HH}). Taking into consideration the equations of state:
\begin{equation}
pV=NkT+\nu W/d\mbox{ and }K=dNkT/2,
\end{equation}
one can repeat the same procedure presented in subsection \ref{sub.Gibbs-Duhem} to obtain the corresponding Gibbs-Duhem relation:
\begin{equation}\label{exp.Gibbs-Duhem}
S=\frac{1}{T}\left(\kappa U+\gamma pV-\mu N\right).
\end{equation}
The coefficients $\kappa$ and $\gamma$ depends on $(d,\nu)$ as:
\begin{equation}\label{scaling.exp}
\kappa=\frac{4d-d\nu-2\nu}{d(2-\nu)}\mbox{ and }\gamma=-\frac{d-2}{2-\nu},
\end{equation}
which are the exponents of the following scaling laws:
\begin{equation}\label{gen.scaling}
S\left(\alpha^{\kappa}U,\alpha^{\gamma}V,\alpha N\right)=\alpha S(U,V,N).
\end{equation}
According to the present results, the generic Hamiltonian system (\ref{generic.HH}) should obey the following thermodynamic limit $N\rightarrow+\infty$:
\begin{equation}
U/N^{\kappa}=const\mbox{ and }V/N^{\gamma}=const.
\end{equation}
In particular, the results (\ref{non.extensive.th}), (\ref{Duhem-Gibbs}) and (\ref{scaling.sgg}) with exponents $\kappa=7/3$ and $\gamma=-1$ are obtained for $d=3$ and $\nu=1$. However, the identity (\ref{exp.Gibbs-Duhem}) undergoes the same limitation of its extensive counterpart: it is a non-universal relation. In other words, the concrete mathematical form of this type of thermodynamical identity should depend on the specific \emph{scaling properties} of the entropy $S(U,V,N)$, \emph{whose existence is not a priori guaranteed for any non-extensive system}. Anyway, it is important to remark that this perspective does not deny, but complements the one proposed by Latella and co-workers.

A model system with nontrivial scalings law is the classical self-gravitating gas of identical non-relativistic particles with a hard-core radius $a$, which regularizes the short-range divergence of Newtonian inter-particles interaction energy as follows:
\begin{equation}
w\left(\mathbf{r}_{i},\mathbf{r}_{i}|a\right)=\left\{
\begin{array}{cc}
  +\infty & \mbox{ if }\left|\mathbf{r}_{i}-\mathbf{r}_{j}\right|<a, \\
  -Gm_{i}m_{j}/\left|\mathbf{r}_{i}-\mathbf{r}_{j}\right| & \mbox{ otherwise.}
\end{array}
\right.
\end{equation}
The finite size of system particles can be disregarded in the limit of large energies and low densities, where one should expect the applicability of the non-extensive thermodynamic limit (\ref{non.extensive.th}). However, the non-vanishing hard-core radius $a$ turns relevant in the limit of low energies and high densities, where one should expect the applicability of the non-extensive thermodynamic limit (\ref{cerruti.tl}) introduced by Cerruti-Sola et al \cite{Cerruti}. The concrete form of scaling laws of this self-gravitating system outside these two limit situations, if exist, does not follow the power-law behaviors as the one shown in equation (\ref{gen.scaling}).

\subsection{Merge of two identical astrophysical systems}

Let us finally analyze the incidence of nonextensivity on the merge of astrophysical systems. This is an example of an out-of-equilibrium process. During the relaxation dynamics that follows after the merge, a fraction of particles can escape from the compound system. For the sake of simplicity, let us suppose that the number of these evaporation events is very small, which enable us to consider that the total energy and the number of particles remain approximately fixed during the relaxation. Of course, this is a very simplifying assumption whose validity should be verified from their consequences. Let us consider here two different situations: (i) the merge of two identical astrophysical systems that are isolated from the gravitational influence of other astrophysical systems, and (ii) the merge of two identical astrophysical systems that are under the gravitational influence of a third very large astrophysical system.

For the first situation, let us employ the estimation (\ref{mic.entropy}) considering two identical isolated astrophysical systems with energy $U$ and particles number $N$. Let us suppose that these systems were infinitely far each other at the initial state, so that, their interaction energy is exactly zero and their total entropy is additive and given by $S_{i} = 2S\left(U,N\right)$. Once the compound system finishes its relaxation, its final entropy can be evaluated from the entropy (\ref{mic.entropy}) with energy $U_{f} = 2U$ and particles number $N_{f} = 2N$, and therefore:
\begin{equation}
S_{f} = S\left(2U,2N\right) = 2S\left(U,N\right) + 4Nk\log2.
\end{equation}
Since the final entropy $S_{f}$ is greater than the initial entropy $S_{i}$, the merge of the astrophysical systems is an \emph{irreversible process} that leads to increase thermodynamic stability. This situation differs from the one considered in \emph{Gibbs paradox} \cite{Reichl}, where the total entropy remains unchanged after the merge of two identical extensive systems (e.g.: gases or liquids initially divided by a separation wall). Temperature does not change during the merge, $kT\sim -2U/3N$. However, the characteristic linear size of the system $R_{c}\sim GM^{2}/\left|U\right|$ is duplicated. It means that the merge leads to a reconfiguration of particles distribution. Hence, the growth of the entropy is related to the fact that the system final volume $V_{f}=8V$ is four times the initial volume $V_{i}=2V$. This behavior is a consequence of non-extensive character of astrophysical systems.

Let us now consider the second situation: the merge of two identical open astrophysical systems (e.g.: two globular clusters) that are under the gravitational influence of a third very large astrophysical system (e.g.: their host galaxy). Formally, each system of this situation could be described by King model discussed in subsection \ref{King.section}. Let us suppose that these systems evolve in opposite directions along a same circular orbit of around a galaxy with a very large mass $M_{G}\gg M_{s}$. This consideration enables us neglect the interaction energy $W$ between them in comparison with their respective internal energies $U_{s}$ when they are located in diametral position in that circular orbit. Precisely, the internal energy of these systems is of order $\left|U_{s}\right|\sim GM^{2}_{s}/R_{s}$, while the interaction energy $\left| W\right|\sim GM^{2}_{s}/2R_{O}$, where $R_{s}\sim R_{O}(M_{s}/2M_{G})^{1/3}$ (tidal radius of equation (\ref{tidal.quad}) above). Since $ M_{s}\ll M_{G}\Rightarrow \left|W\right|\ll \left|U\right|$. After the formation of compound system and its relaxation, the final values of the total energy and the mass of the compound system (c) duplicate the initial values of each system, $M_{c}=2M_{c}$ and $U_{c}=2U_{s}$, while its radius only increases as $R_{c}=2^{1/3}R_{s}$. Accordingly, the final dimensionless energy $u_{c}=2^{-2/3}u_{s}$ is now increased. Note that $u_{c}>u_{s}$ because of $u_{s}<0$ (for bounded stable configurations). Once again, the total entropy of the compound system is greater than the one of the initial configuration:
\begin{equation}
\Delta S=S_{c}\left(U_{c},M_{c},R_{c}\right)-2S\left(U_{s},M_{s},R_{s}\right)=2N\left[s(u_{c})-s(u_{s})\right]+2N\ln 2>0.
\end{equation}
The difference of the reduced entropy $s(u_{c})-s(u_{s})$ is positive since $\eta(u)=\partial s(u)/\partial u>0$, with $\eta(u)=GMm/TR_{t}$ is the dimensionless inverse temperature. This difference can be evaluated from the integral:
\begin{equation}
s(u_{c})-s(u_{s})=\int^{u_{c}}_{u_{s}}\eta(u)du>0
\end{equation}
using the temperature \emph{versus} energy dependence shown in figure \ref{King.eps}. The additional contribution $2N\ln 2$ was obtained assuming the relevance of the factor $\sim3N\ln\left[R_{t} N^{1/3}\right]$ in analogy with expression (\ref{additional.term}), which considers the scaling dependence of the characteristic linear size for a non-relativistic gas of identical point particles. Interestingly, the present situation should also obey an additional restriction: the final configuration of the compound system could be stable if its dimensionless energy $u_{c}$ is lower than the critical energy of the \emph{evaporative disruption} $u_{C}=-0$.$601$. Accordingly, the final configuration of the compound system will be \emph{unstable} if the initial energy $u_{s}$ of each system belongs to the interval $u^{*}_{C}<u_{s}<u_{C}$ with $u^{*}_{C}=2^{2/3}u_{C}=-0$.$954$, and therefore, it will undergo a sudden evaporation to release its excess of energy and particles. The initial hypothesis about the constancy of the total energy and the number of particles cannot be fulfilled in this case. This is other example that an external influence can introduce a great affectation in some thermodynamic behaviors and processes in astrophysics.

\section{Final remarks}

Conventional thermo-statistical treatments and some of their associated notions have a restricted applicability in the context of astrophysical systems. The long-range character of gravitation presupposes a strong limitation for the licitness of notions such as separability and additivity, thermal contact, extensive and intensive quantities, and hence, applicability of conventional statistical ensembles. So far, the usage of these notions and frameworks is highly motivated by our everyday practical experience with large extensive systems. In one way or another, these idealizations play a crucial role in our understanding of macroscopic phenomena of this context, which include thermodynamic equilibrium conditions, transport processes, phase transitions, among others.

Separability and additivity under a thermal contact is not longer a licit idealization to study astrophysical systems in mutual thermodynamic interaction. However, restricted forms of these properties are still possible when there exists an effective separability among the collective and the internal degrees of freedom of these systems. The collective motions are now responsible to drive the interaction among the astrophysical systems. This mechanism replaces the conventional thermal contact and introduces significant modifications in thermal equilibrium conditions [including the stationary conditions (equalization of temperatures) and stability conditions (restrictions over response functions)] and the energy transport. A plausible consequence of this mechanism is that thermal equilibrium could be possible among astrophysical systems with negative heat capacities. Moreover, distribution profiles with very dense cores, which are associated with states with negative heat capacities, favor the stability of collective motions of two interacting astrophysical systems.

Thermodynamic behavior of astrophysical systems exhibits a strong dependence on the external backgrounds conditions. The incidence of a long-range interaction as gravitation imposes that all constituents are strongly correlated among them. In this way, any external influence, even acting on the system boundary, will propagates to the whole astrophysical structure and modifies its macroscopic behavior in a radical way. By itself, this is why the thermodynamics of astrophysical systems is so rich and challenging. Qualitatively speaking, astrophysical systems exhibit general features such as negative heat capacities and gravitational collapse. However, the concrete quantitative characterization of these macroscopic phenomena depends on the external conditions that affect each astrophysical situation.

Although a universal thermodynamic limit for astrophysical systems is not possible, some analogous scaling relations are possible for certain situations. In particular, the scaling laws (\ref{non.extensive.th}) can be regarded as the right thermodynamic limit for a three-dimensional self-gravitating gas of identical non-relativistic point particles regardless their classical or quantum nature. This thermodynamic limit has been already obtained in other works in the literature \cite{Vel.QEM,Chavanis2002,Chavanis2005}, and its relevance is now supported by an astrophysical counterpart of Gibbs-Duhem relation (\ref{Duhem-Gibbs}). For real astrophysical situations, these scaling laws are restricted by the incidence of relativistic effects. According to the arguments presented during comparison with other thermodynamic limits, extensivity of entropy seems to be a more general property than any other criterium to establish a scaling behavior for macroscopic observables (e.g.: the called Kac prescription \cite{Kac}). Conceptually, this idea is quite satisfactory because of the entropy is the central notion of thermo-statistical description.

Several analysis presented in this work are far to be complete. A deeper study of these questions should lead to the new theoretical developments such as the astrophysical counterparts of conventional canonical and gran canonical statistical ensembles, as well as a better understanding of phase transitions and critical phenomena in astrophysics. Our understanding of thermodynamic processes that involve interaction among astrophysical systems still demands much work in future years. A particular attention should be devoted to the peculiarities of the transport processes of energy and matter in presence of long range interactions and systems with negative heat capacities. Conventional kinetic equations that are employed to describe these non-equilibrium processes were not conceived to deal with these macroscopic behaviors.

\section*{Acknowledgments}
Author thanks to Ya\~{n}ez-Valenzuela, Gomez-Leyton, Tello-Ortiz and Morales-Ram\'irez for their suggestions and interesting discussions. Additional thanks to anonymous referees to point out the connections of astrophysical counterpart of Gibbs-Duhem relation with results of Latella and co-workers, as well as other interesting suggestions. Partial financial support from the \textbf{VRIDT-UCN} research program is also acknowledged.

\end{document}